\PassOptionsToPackage{unicode}{hyperref}
\PassOptionsToPackage{hyphens}{url}
\PassOptionsToPackage{dvipsnames,svgnames,x11names}{xcolor}
\documentclass[12pt]{article}

\usepackage{amsmath,amssymb}
\usepackage{iftex}
\ifPDFTeX
  \usepackage[T1]{fontenc}
  \usepackage[utf8]{inputenc}
  \usepackage{textcomp} 
\else 
  \defaultfontfeatures{Scale=MatchLowercase}
  \defaultfontfeatures[\rmfamily]{Ligatures=TeX,Scale=1}
\fi
\usepackage{lmodern}
\ifPDFTeX\else  
\fi
\IfFileExists{upquote.sty}{\usepackage{upquote}}{}
\IfFileExists{microtype.sty}{
  \usepackage[]{microtype}
  \UseMicrotypeSet[protrusion]{basicmath} 
}{}
\makeatletter
\@ifundefined{KOMAClassName}{
  \IfFileExists{parskip.sty}{%
    \usepackage{parskip}
  }{
    \setlength{\parindent}{0pt}
    \setlength{\parskip}{6pt plus 2pt minus 1pt}}
}{
  \KOMAoptions{parskip=half}}
\makeatother
\usepackage{xcolor}
\makeatletter
\ifx\paragraph\undefined\else
  \let\oldparagraph\paragraph
  \renewcommand{\paragraph}{
    \@ifstar
      \xxxParagraphStar
      \xxxParagraphNoStar
  }
  \newcommand{\xxxParagraphStar}[1]{\oldparagraph*{#1}\mbox{}}
  \newcommand{\xxxParagraphNoStar}[1]{\oldparagraph{#1}\mbox{}}
\fi
\ifx\subparagraph\undefined\else
  \let\oldsubparagraph\subparagraph
  \renewcommand{\subparagraph}{
    \@ifstar
      \xxxSubParagraphStar
      \xxxSubParagraphNoStar
  }
  \newcommand{\xxxSubParagraphStar}[1]{\oldsubparagraph*{#1}\mbox{}}
  \newcommand{\xxxSubParagraphNoStar}[1]{\oldsubparagraph{#1}\mbox{}}
\fi
\makeatother

\usepackage{longtable,booktabs,array}
\usepackage{calc} 
\usepackage{etoolbox}
\makeatletter
\patchcmd\longtable{\par}{\if@noskipsec\mbox{}\fi\par}{}{}
\makeatother
\IfFileExists{footnotehyper.sty}{\usepackage{footnotehyper}}{\usepackage{footnote}}
\makesavenoteenv{longtable}
\usepackage{graphicx}
\makeatletter
\def\maxwidth{\ifdim\Gin@nat@width>\linewidth\linewidth\else\Gin@nat@width\fi}
\def\maxheight{\ifdim\Gin@nat@height>\textheight\textheight\else\Gin@nat@height\fi}
\makeatother
\setkeys{Gin}{width=\maxwidth,height=\maxheight,keepaspectratio}
\makeatletter
\def\fps@figure{htbp}
\makeatother

\usepackage{xr-hyper} 
\usepackage{xr}
\makeatletter
\@ifpackageloaded{caption}{}{\usepackage{caption}}
\AtBeginDocument{%
\ifdefined\contentsname
  \renewcommand*\contentsname{Table of contents}
\else
  \newcommand\contentsname{Table of contents}
\fi
\ifdefined\listfigurename
  \renewcommand*\listfigurename{List of Figures}
\else
  \newcommand\listfigurename{List of Figures}
\fi
\ifdefined\listtablename
  \renewcommand*\listtablename{List of Tables}
\else
  \newcommand\listtablename{List of Tables}
\fi
\ifdefined\figurename
  \renewcommand*\figurename{Figure}
\else
  \newcommand\figurename{Figure}
\fi
\ifdefined\tablename
  \renewcommand*\tablename{Table}
\else
  \newcommand\tablename{Table}
\fi
}
\@ifpackageloaded{float}{}{\usepackage{float}}
\floatstyle{ruled}
\@ifundefined{c@chapter}{\newfloat{codelisting}{h}{lop}}{\newfloat{codelisting}{h}{lop}[chapter]}
\floatname{codelisting}{Listing}

\makeatother
\makeatletter
\@ifpackageloaded{caption}{}{\usepackage{caption}}
\@ifpackageloaded{subcaption}{}{\usepackage{subcaption}}
\makeatother

\ifLuaTeX
  \usepackage{selnolig}  
\fi
\usepackage[]{natbib}
\usepackage{bookmark}
\usepackage{amsthm}
\usepackage{tabularx}
\usepackage{makecell}
\usepackage{array}

\IfFileExists{xurl.sty}{\usepackage{xurl}}{} 
\hypersetup{
  pdftitle={Title},
  pdfauthor={Author 1; Author 2},
  pdfkeywords={3 to 6 keywords, that do not appear in the title},
  colorlinks=true,
  linkcolor={blue},
  filecolor={Maroon},
  citecolor={Blue},
  urlcolor={Blue},
  pdfcreator={LaTeX via pandoc}}

\newcommand\independent{\protect\mathpalette{\protect\independenT}{\perp}}
\def\independenT#1#2{\mathrel{\rlap{$#1#2$}\mkern2mu{#1#2}}}
\RequirePackage{bm}

\newtheorem{prop}{Proposition}

\newtheorem{remark}{Remark}[section]
\newcommand{\Mean}{{\mathbb{E}}}

\def\mB{\mathcal{B}}
\def\mY{\mathcal{Y}}
\def\mE{\mathcal{E}}

\newcommand{\anon}{1}

\begin{document}

\def\spacingset#1{\renewcommand{\baselinestretch}%
{#1}\small\normalsize} \spacingset{1}


\if1\anon
{
  \title{\bf How much does Home Field Advantage matter in Soccer Games? A causal inference approach for English Premier League analysis}

    \author{Chen~Wang\textsuperscript{1}\thanks{tiachenw8@connect.hku.hk},
    Katherine~Price\textsuperscript{2}\thanks{pricekm@mail.missouri.edu}, 
    Hengrui~Cai\textsuperscript{3}\thanks{hengrc1@uci.com},\\
    Weining~Shen\textsuperscript{3}\thanks{weinings@uci.edu}, 
    Zhanrui~Cai\textsuperscript{1}\thanks{zhanruic@hku.hk},
    and Guanyu~Hu\textsuperscript{4}\thanks{huguanyu@msu.edu}\\
    \\
\textsuperscript{1}Faculty of Business and Economics, The University of Hong Kong\\
\textsuperscript{2}Department of Statistics, University of Missouri, Columbia\\
\textsuperscript{3}Department of Statistics, University of California, Irvine\\
\textsuperscript{4}Department of Statistics and Probability, Michigan State University
}
  \maketitle
} \fi

\if0\anon
{
  \bigskip
  \bigskip
  \bigskip
  \begin{center}
    {\LARGE\bf Title}
\end{center}
  \medskip
} \fi

\bigskip
\begin{abstract}
In many sports, it is commonly believed that the home team has an advantage over the visiting team, known as the home field advantage. Yet its causal effect on team performance is largely unknown. In this paper, we propose a novel causal inference approach to study the causal effect of home field advantage in English Premier League. We develop a hierarchical causal model and show that both league level and team level causal effects are identifiable and can be conveniently estimated. We further develop an inference procedure for the proposed estimators and demonstrate its excellent numerical performance via simulation studies. We implement our method on the 2020-21 English Premier League data and assess the causal effect of home advantage on eleven summary statistics that measure the offensive and defensive performance and referee-assessed disciplinary outcomes. We find that the home field advantage resides more heavily in offensive statistics than it does in defensive or referee statistics. We also find evidence that teams that had lower rankings retain a higher home field advantage.
\end{abstract}

\noindent%
\vfill

\newpage
\spacingset{1.8} 

\section{Introduction}\label{sec-intro}
Quantitative analysis in sports analytics has received a great deal of attention in recent years. Traditional statistical research for sports analytics mainly focused on game result prediction, such as predicting the number of goals scored in soccer matches \citep{dixon1997modelling, karlis2003analysis,baio2010bayesian}, and the basketball game outcomes \citep{carlin1996improved,caudill2003predicting,cattelan2013dynamic}. More recently, fast development in game tracking technologies has greatly improved the quality and variety of collected data sources \citep{albert2017handbook}, and in turn substantially expanded the role of statistics in sports analytics, including performance evaluation of players and teams  \citep{cervone2014pointwise,franks2015characterizing,wu2018modeling,hu2020bayesian,hu2021cjs,yin2023analysis,qi2024made}, commentator's in-game analysis and coach's decision making \citep{fernandez2018wide,sandholtz2019measuring,grieshop2025continuous}. 

The home field advantage is a key concept that has been studied across different sports for over a century. The first public attempt to statistically quantify the home field advantage was about 40 years ago \citep{physicaleducation}, where \citet{schwartz1977home} studied the existence of home advantage in several sports and found that its effect was most pronounced in the indoor sports such as ice hockey and basketball. 
Since then, statisticians and psychologists have committed themselves to mapping out this anomaly, and there is some consensus on it. The percentages found through different studies in general float around 55-60\% of an advantage to the home team (wherein 50\% would mean there is no advantage either way). \citet{courneya1992home} found that different sports in general held different advantage percentages: 57.3\% for American football versus 69\% for soccer, for example. In a study of basketball home court advantage, \citet{calleja2018brief} found that the visiting team scored 2.8 points less, attempted 2 fewer free throws, and made 1 less free throw than their season averages. Season length is also shown to have a negative correlation with the strength of home field advantage, where sports with 100 games or more per season were shown to have significantly less advantage than a sport with fewer than 50 games per season \citep{Metaanalysis}. This finding can be explained by the fact that each game becomes less important on average as the number of games increases. Some studies have shown that the home field advantage becomes less significant in the modern era. For example, \citet{Metaanalysis} found that the home field advantage was the strongest in the pre-1950s, as opposed to any other twenty-year block leading up to 2007, the year he conducted the study. In another work, \citet{goumas2017modelling} showed that between 2003 and 2013, the home field advantage for the UEFA Champions League decreased by 1.8\%, supporting the claim that as time goes on, the home field advantage decreases to a certain level and remains still. In general, the home field advantage tends to deteriorate as the athletes play longer on an away field. 
	
In principle, there are four major factors associated with the home field advantage: crowd involvement, travel fatigue, familiarity of facilities, and referee bias that benefits the home teams. Several studies \citep{schwartz1977home,greer1983spectator,benz2024comprehensive} have shown a positive correlation between the size of the crowd and the effect of the home field advantage, an advantage that can get as high as 12\% over a team's opponents. Travel fatigue is another important factor for home field advantage \citep{Metaanalysis,calleja2018brief,NFLtimezone}, since the athletes have an overall reduction in mean wellness, including a reduction of sleep, self-reported feelings of jet lag and energy reduction; the importance of sleep and proper recovery cannot be over-exaggerated, and it can be difficult to do either of those properly in unfamiliar settings (e.g., a hotel room or a bus). Furthermore, familiarity of facilities gives the home team a strong advantage over a visiting team \citep{MovingNewStadium}. A home team not only does not have to travel to an unfamiliar facility, be away from home, and spend hours transporting themselves; they also get to use their own facility, locker room, and field. It cannot be underestimated the value of knowing and knowing well the nooks and crannies of a facility. Ultimately, giving rule advantages to the home team is very common in certain sports \citep{courneya1992home}. For example, in baseball the home team gets the advantage of batting last, giving them the final opportunity to score a run in the game. In hockey, the last line change goes to the home team.

Soccer has proven to have one of the highest home field advantages among major sports in a majority of current research. \citet{MovingNewStadium} states that the home field advantage in soccer is equivalent to 0.6 goals per game and the visual cues that come from being intimately familiar with a facility can be exponentially helpful in fast paced sports such as soccer, which could perhaps explain why the home field advantage is less pronounced in slower, stopping sports such as baseball. \citet{sanchez2009analysis} observes that the better a soccer team is, the more often the home field advantage appears, after studying home field advantage across groups of variably ranked soccer teams. 

Despite the vast amount of existing work on home field advantage quantification in soccer and other sports \citep{leitner2020no,benz2021estimating,fischer2021does}, very little is known about its {\it causal effect} on team performance; and it is our goal in this paper to fill this gap. Existing studies typically rely on associational analyses without accounting for the structural dependencies in match assignments—specifically, the fact that one team's home status necessarily implies the other's away status. This lack of independent treatment assignment introduces a form of confounding or interference, making it difficult to isolate the true causal effect of playing at home from correlated team or contextual factors.
 In particular, we study soccer games by analyzing a data set collected from the English Premier League (EPL) 2020-2021 season, where 380 games were played, that is, two games between each pair of 20 teams. We choose eleven team-level summary statistics as the main outcomes that represent the team's performance on defensive, offensive, and referee sides. We then develop a new causal inference approach for assessing the causal effect of home field advantage on these outcomes. More details about our data application are provided in Section \ref{sec:data}. 

Our work complements and contrasts with that of \citet{lopez2018often}, who developed a unified Bayesian framework to quantify the role of randomness versus team quality across North American sports leagues. While their paper does not include soccer, it provides valuable insights into how often the best team wins and includes important findings on home field advantage across major sports in north America. They show that HFA varies by sport and team, with implications for league fairness. However, their approach is primarily simulation-based and centered on outcome probabilities at the season or championship level. In contrast, our study focuses on match-level causal effects using a potential outcomes framework, offering causally identified insights into how HFA shapes specific aspects of team behavior, particularly offensive dynamics. Together, these perspectives help build a more complete understanding of fairness and structure in competitive sports.

In causal inference literature (for observational studies), causal effect estimation of a binary treatment is a classic topic that has been intensively studied \citep{licausal,imbens2004nonparametric}. 
Over the past few decades, a number of methods in both statistics and econometrics have been proposed to identify and estimate the average treatment effects of a particular event/treatment \citep[see a recent overview in][]{athey2017estimating} under the assumption of ignorability (also unknown as unconfounded treatment assignment) \citep{rosenbaum1983central}, including regression imputation, (augmented) inverse probability weighting \citep[see e.g., ][]{horvitz1952generalization, rosenbaum1983central, robins1994estimation, bang2005doubly, cao2009improving}, and matching \citep[see e.g., ][]{rubin1973matching, rosenbaum1989optimal, heckman1997matching, hansen2004full, rubin2006matched, abadie2006large, abadie2016matching}. However, the data structure in soccer games is distinguished from those in the existing literature, leading to {\it two unique challenges} in identifying the causal effects of home field advantage. First, the EPL data set is obtained neither from a randomized trial or observational studies as standard in causal inference literature, but instead is based on a collection of pair-wise matches between every two teams. For one season, each team will have one home game and one away game with each of the other 19 opponents. All matches are pre-scheduled. Thus, propensity-score-based methods can hardly gain efficiency with such a design. Secondly, for each match, there is one team with home field advantage and the other without, i.e., there are no matches in the neutral field. In fact, this is the common practice for other major professional sports leagues such as NBA, NFL, and MLB. In other words, there is technically no control group where both teams have no home field advantage. Hence, we cannot rely on the matching-based method to estimate the causal effects. To overcome these difficulties, in this paper, we establish a  hierarchical causal model to characterize the underlying true causal effects at the league and team levels, and propose a novel causal estimation approach for home field advantages with inference procedures.

Our proposed method is unique in the following aspects. First, the idea of {\it pairing home and away games} for solving causal inference problems is novel. In fact, this idea and our proposed approach are widely applicable to general sports applications such as football, baseball, and basketball studies, and provide a valuable alternative to the existing literature that mainly relies on propensity score. Secondly, under the proposed hierarchical model framework, both league level and team level causal effects are identifiable and can be conveniently estimated. Moreover, our inference procedure is developed based on linear model theory that is accessible to a wide audience including first-year graduate students in statistics. Thirdly, our real data analysis results reveal several interesting findings from England Primer League, which may provide new insights to practitioners in the sports industry. It is our hope that the data application presented in this paper as well as the developed statistical methodology can reach a wide range of audience, be useful for educational purposes in statistics and data science classes, and stimulate new ideas for more causal analysis in sports analytics. 

The rest of the paper is organized as follows. In Section \ref{sec:data}, we
give an overview of our motivating data application and introduce several representative summary outcomes. In Section~\ref{sec:model}, we introduce the causal inference framework and propose hierarchical causal models for both team-level and league-level home field advantage effect estimation under univaraite and multivaraite settings. Extensive simulation studies are presented in Section~\ref{sec:simu} to investigate the empirical performance of our approach. 
We apply our method to analyze the 2020-2021 England Primer League in Section~\ref{sec:app} and conclude with a discussion of future directions in Section~\ref{sec:disc}.

\section{Motivating Data}\label{sec:data}
We are interested in studying the causal effect of home field advantage for in-game performance during English Premier League 2020-2021 season. Despite as obvious as it sounds, the home field advantage is in fact not evident at first glance through the usual descriptive statistics such as the game outcomes and number of scored goals. For example, among the 380 games in that season, the number of home wins was 144 (37.89\%), which is even less than the number of away wins, 153 (40.26\%). Among the total of 1024 goals scored in that season, 514 (50.19\%) were scored by the home team, and 510 (49.81\%) by the away team. 

To better understand this phenomenon, we study a data set provided by Hudl \& Wyscout, a company that excels at soccer game scouting and match analysis. The data is collected from 380 games played by 20 teams in EPL, and includes 
a number of statistics collected from each game that range across the defensive and offensive capabilities of the home and away teams and players. The data that support the findings of this study are available from Hudl \& Wyscout. Restrictions apply to the availability of these data, which were used under license for this study. Data are available from the corresponding author with the permission of Hudl \& Wyscout (\url{www.hudl.com}). 
We choose to focus on eleven in-game statistics as follows:
\begin{itemize}
    \item \textbf{Attacks w/ Shot} - The number of times that an offensive team makes a forward move towards their goal (a dribble, a pass, etc) followed by a shot;
    \item \textbf{Defence Interceptions} - The number of times that a defending team intercepts a pass; 
    \item \textbf{Reaching Opponent Box} - The number of times that a team moves the ball into their opponent's goal box;
    \item \textbf{Reaching Opponent Half} - The number of times that a team moves the ball into their opponent's half;
    \item \textbf{Shots Blocked} -  The number of times that a defending team deflects a shot on goal to prevent scoring;
    \item \textbf{Shots from Box} - The number of shots taken from the goal box;
    \item \textbf{Shots from Danger Zone} -  The number of shots taken from the "Danger Zone", which is a relative area in the center of the field approximately 18 yards or less from the goal; 
    \item \textbf{Successful Key Passes} - The number of passes that would have resulted in assists if the resulting shot had been made;
    \item \textbf{Touches in Box} - Number of  passes or touches that occur within the penalty area;
    \item \textbf{Successful Key Passes} - The number of passes that would have resulted in assists if the resulting shot had been made;
    \item \textbf{Expected Goals (XG)} - The average likelihood a goal will be scored given the position of the player over the course of a game; 
    \item \textbf{Yellow Cards} - The number of yellow cards given to a team in a game.  
\end{itemize}

Table \ref{tab:summary_statistics} contains the eleven summary statistics from the raw data, where each row shows the statistic, its primary role in the game (defense, offense, or referee), the means for the home team and for the away team, respectively, and the overall standard deviation. These in-game statistics are chosen because they are most relevant to studying the home field advantage and also sufficient to cover different aspects of soccer games (e.g., team offensive and defensive performance). 
\begin{table}[th]
    \centering
        \caption{Selective summary statistics for 2020-2021 England Primer League.}
    \label{tab:summary_statistics}
    \begin{tabular}{  c  c c  c  c  } 
  \toprule
  \textbf{Statistic} & \textbf{Role} & \textbf{Mean Home} & \textbf{Mean Away} & \textbf{Overall SD}\\

  \midrule
   Attacks w/ Shot & Offense & 11.547 & 9.979 & 4.910 \\ 

  Defence Interceptions & Defense & 42.639 & 44.637 & 11.287  \\

  Reaching Opponent Box & Offense &  14.779 & 13.097 & 6.135\\

  Reaching Opponent Half & Offense & 58.561 & 54.968 &  12.566\\

  Shots Blocked & Defense & 3.382 & 2.95 & 2.273 \\

  Shots from Box & Offense & 7.5 & 6.434 & 3.594 \\

  Shots from Danger Zone & Offense & 5.058 & 4.271 &  2.622\\

  Successful Key Passes & Offense & 3.584 & 3.037 & 2.328\\

  Touches in Box & Offense & 19.468 & 17.216 & 8.883 \\

  Expected Goals (XG) & Offense & 1.577 & 1.395 & 0.867\\

  Yellow Cards & Referee & 1.447 & 1.474 & 1.151\\
  \bottomrule
\end{tabular}

\end{table}

\begin{figure}
    \centering
   \includegraphics[width=1\textwidth]{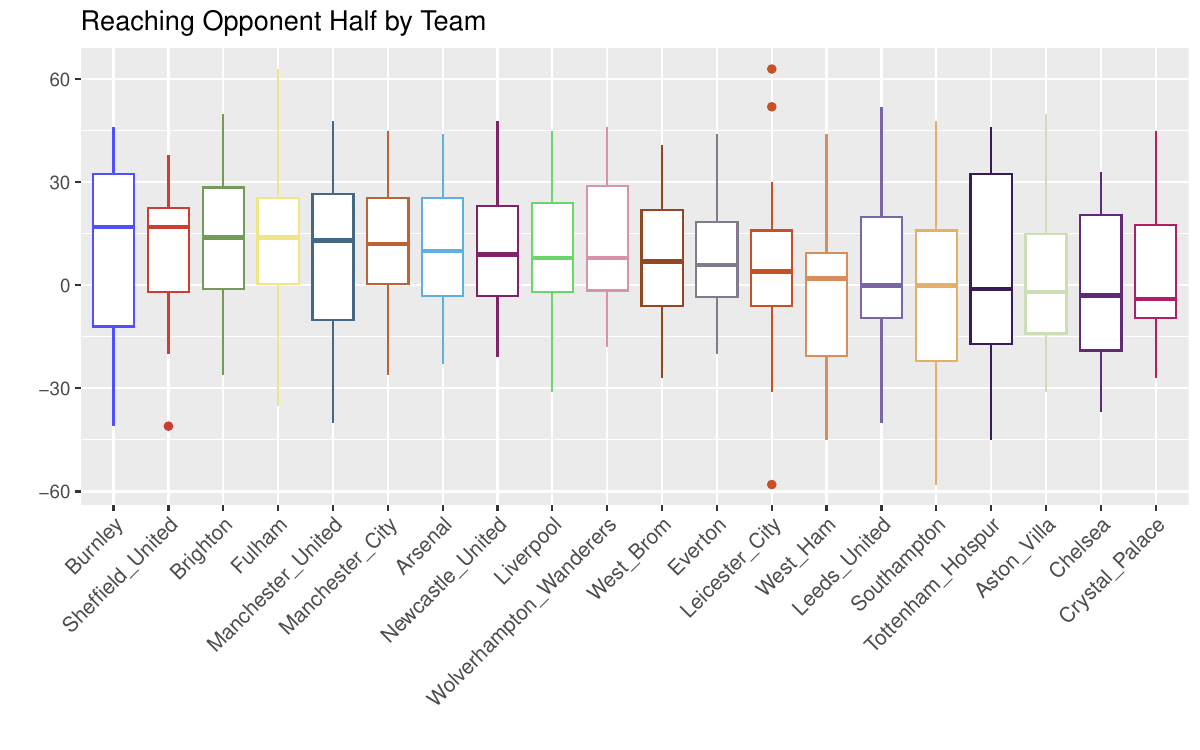}
    \caption{Net difference of Reaching Opponent Half for 20 teams.}
    \label{fig:reaching}
\end{figure}
\begin{figure}
    \centering
   \includegraphics[width=1\textwidth]{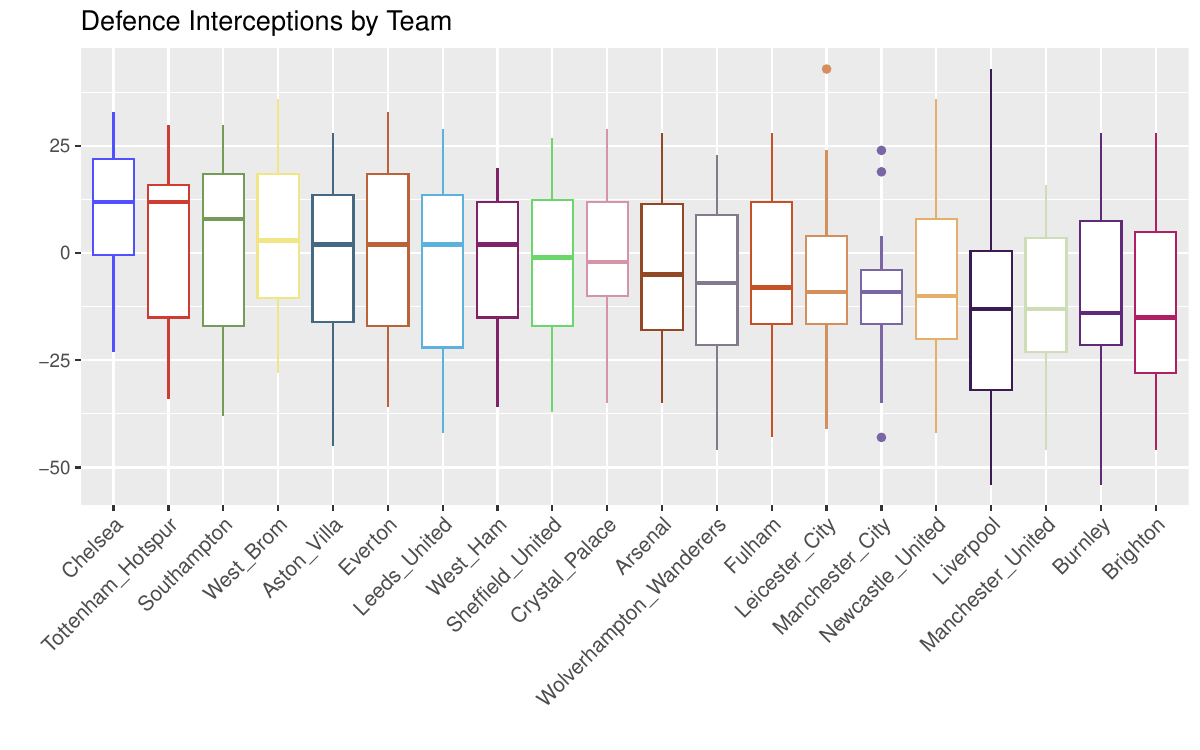}
    \caption{Net difference of Defence Interceptions for 20 teams.}
    \label{fig:Interception}
\end{figure}
\begin{figure}
    \centering
   \includegraphics[width=1\textwidth]{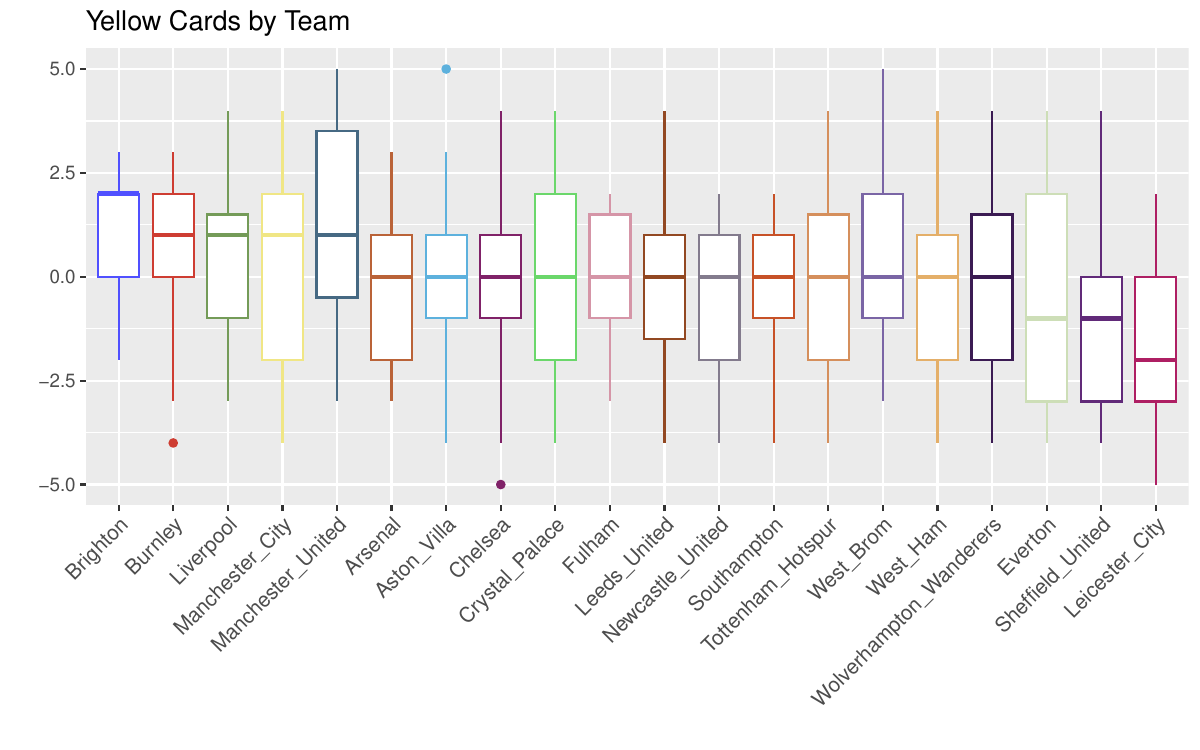}
    \caption{Net difference of Yellow Cards for 20 teams.}
    \label{fig:yellow}
\end{figure}

To better understand the role of these selected statistics. We choose three of them (representing offense, defense, and referee), calculate the difference in these statistics between home and away games for each team, and present their distributions in Figure~\ref{fig:reaching} to Figure~\ref{fig:yellow}. 
The offensive statistic is Reaching Opponent Half, and its distribution for different teams is illustrated in Figure \ref{fig:reaching}. The difference between home and away teams, in theory, should be positive overall if the home field advantage is present, negative if there is actually an advantage towards the away team, and 0 if there is no advantage either way. From the picture, we can see a clear trend for home advantage, e.g., 16 of the 20 teams have a positive mean, which implies that they reach their opponents' half more often on average when they play at home as opposed to away. This finding is in fact fairly consistent across all offensive variables. The four teams that do not show evidence of home field advantage for Reaching the Opponent's Half are Tottenham Hotspur, Aston Villa, Chelsea, and Crystal Palace. The top three teams from that season (Manchester City, Manchester United, and Liverpool) exhibit similar patterns in the picture.

 The Defence Interceptions are chosen as a representation for the defensive statistic; and we present its distribution of the difference between home and away teams in Figure \ref{fig:Interception}. Since defence interceptions negatively impact a team, the home field advantage will be seen here if, oppositely to the offensive statistic, the distribution of the difference is skewed negatively. Eight teams do not show evidence of home field advantage: Chelsea, Tottenham Hotspur, Southampton, West Bromwich Albion, Aston Villa, Everton, Leeds United, and West Ham United. Interestingly, the top three teams from that season (Manchester City, Manchester United, and Liverpool) all have negative means but show a different level of variability in this statistic. 

Finally, referee-assessed disciplinary outcome is represented by the number of yellow cards at the team level in each match. The difference between a team's yellow card calls at home versus away is shown in Figure~\ref{fig:yellow}. If there was a home field advantage effect in yellow cards given, the boxplot would trend negatively. Different from the offensive and defensive statistics, we see a majority of teams centralizing their distributions around the zero line, which is an indication that the referee calls are similar whether the team is home or away. Out of 20 teams, 12 have means of about zero, and the other 8 have a mix of positive and negative means.

\section{Method}\label{sec:model}
\subsection{Framework and Assumptions}
Suppose there are $n$ different teams in the league, denoted by $\{T_1,\cdots, T_n\}$, with a total number of $N\equiv n(n-1)$ matches between any pair of two teams, $T_i$ and $T_j$, where $i\not = j$ and $i,j \in \{1,\cdots,n\}$. Without loss of generality, we assume $n\geq 3$. 
Therefore, each team $T_i$ has $(n-1)$ matches with home field advantage and another $(n-1)$ without. 
Define a treatment indicator $\delta_{i}=1$ if team $T_i$ has the home field advantage in a match against team $T_j$ for $j\not = i$, and $\delta_{i}=0$ otherwise. 
By definition, we have $\delta_{i} + \delta_{j} =1$ for any match between $T_i$ and $T_j$. 
Consider $K$ in-game summary statistics. For the $k$-th statistic, the main outcome is defined as the net difference in the outcome of interest, i.e., $Y_{i,j}^k =(c_i^k - c_j^k)\in\mathbb{R}$, where $c_i^k$ and $c_j^k$ are the $k$-th statistic described in Section \ref{sec:data} for team $T_i$ 
and team $T_j$, $k = 1, 2, \dots, 11$. 
In Section 3.1 and 3.2, we model each in-game summary statistic separately. Thus we fix $k$ and will drop $k$ in the notation. Following the potential outcome framework \citep[see e.g., ][]{rubin1974estimating}, we define the potential outcome $Y_*(\delta_{i}=a,\delta_{j}=1-a)$ as the outcome of interest that would be observed after the match between team $T_i$ and team $T_j$, where $a=1$ or $a=0$ corresponds to that team $T_i$ or team $T_j$ has the home field advantage, respectively. We define the baseline matchup strength between teams \(T_i\) and \(T_j\) on a neutral field as \(\alpha_{i,j} = \mathbb{E}\{Y^*(\delta_i = 0, \delta_j = 0)\}\), representing the expected net outcome when neither team holds home field advantage. The potential outcome \(Y^*(\delta_i = 1, \delta_j = 1)\) is not well-defined, as only one team can be assigned home field status in a given match; thus, it is excluded from our framework.
As standard in the causal inference literature \citep[see e.g., ][]{rosenbaum1983central}, we make the following assumptions for any pairs of $i \neq j$. 


\noindent \textbf{(A1)}. Stable Unit Treatment Value Assumption: 
\begin{equation*}
\begin{split}
Y_{i,j}= \delta_{i}(1- \delta_{j})Y_*(\delta_{i}=1,\delta_{j}=0)  +(1- \delta_{i}) \delta_{j} Y_*(\delta_{i}=0,\delta_{j}=1).
\end{split}
\end{equation*}

\noindent \textbf{(A2)}. Ignorability: 
\begin{equation*} 
\{Y^*(\delta_{i}=1,\delta_{j}=0),Y^*(\delta_{i}=0,\delta_{j}=1)\}\independent \{\delta_{i},\delta_{j}\}. 
\end{equation*}

\noindent \textbf{(A3)}. Non-monotonicity: If team A dominates team B and team B dominates team C, team C can still dominate team A with a certain probability.

Assumptions (A1) and (A2) are standard in causal inference literature \citep[see e.g.,][]{athey2017estimating}, to ensure that the causal effects are estimable from observed data. By game design, each team will compete with the rest teams with and without home field advantage once, respectively, hence the ignorability assumption holds automatically in our study. Assumption (A3) is to rule out the correlation between different matches that involve the same team. In reality, there are style rivalries between teams in soccer games. Hence the transitive relation does not always hold. 

\subsection{Hierarchical Causal Modeling and Estimation}
 In this section, we detail the proposed hierarchical causal model and its estimation and inference procedures. Specifically, we are interested in team level and league level estimators of
causal effects. To this end, we define the causal effect of home field advantage associated with team $T_i$ as
\begin{eqnarray}\label{def_betai}
\beta_i = \Mean_j \{Y^*(\delta_{i}=1,\delta_{j}=0)-Y^*(\delta_{i}=0,\delta_{j}=0)\},~~\text{for}~i=1,\ldots,n,
\end{eqnarray}
and the causal effect of home field advantage for the entire league as
\begin{eqnarray}\label{def_delta}
\Delta = \Mean_i \{\beta_i\}.
\end{eqnarray}
Given a finite number of teams in the league, we are interested in two estimands: the team-specific home field advantage $\hat\beta_i$, and the average home field advantage of the league $\hat\Delta=\sum_{i=1}^n \hat\beta_i /n $ and 
Yet, since there is no match in the neutral field in major professional sports, we always have $\delta_{i} + \delta_{j} =1$, i.e.,  $Y^*(\delta_{i}=0,\delta_{j}=0)$ can never be observed. To address this difficulty and estimate $\beta_i$ from the observational studies, we propose to decompose the outcome function into two parts, one corresponding to the home field advantage and the other representing the potential outcome in a hypothetical neutral field, via a mixed two-way ANOVA design. To be specific, we assume that the outcome of a match between team $T_i$ (with home field advantage, $\delta_i=1$) and team $T_j$ (without home field advantage, $\delta_j=0$) can be modeled parametrically by
\begin{eqnarray}\label{main_model_beta} 
 Y_{i,j}=\alpha_{i,j}+ \beta_{i}+\epsilon_{i,j},
\end{eqnarray}
where $\alpha_{i,j}= \Mean \{Y^*(\delta_{i}=0,\delta_{j}=0)\}$ is the expected net outcome between team $T_i$ and $T_j$ in a hypothetical neutral field (i.e., if there is no team taking home field advantage) based on Assumption (A3), 
and $\epsilon_{i,j}$ is random noise with $\mathcal{N}(0,\sigma^2_0)$. The factorization in \eqref{main_model_beta} enables us to unravel the  home field advantage at the team level by utilizing the pair-wise match design in soccer games that we will discuss shortly. Assumption (A3) is required for \eqref{main_model_beta} so that we can allow independent baseline effect $\alpha_{i,j}$ for any pair of two teams in a hypothetical neutral field. We note that the decomposition in Equation~\eqref{main_model_beta} relies on the assumption that the homefield advantage $\beta_i$ operates additively and independently of the baseline pairwise strength $\alpha_{i,j}$. If, in reality, the homefield effect interacts with the relative strength of the teams, then $\beta_i$ would not be constant across opponents, and the estimand defined by Equation~\eqref{main_model_beta} may be biased. To deal with such a circumstance, our parametric model can be extended by including interaction terms that captures pair-specific deviations from the additive structure, or by specifying a random-coefficient formulation for $\beta_i$ that varies with team strength.
In addition, we adopt a hierarchical model \citep{berry2013bayesian,chu2018bayesian,geng2020mixture} to characterize the relationship between the home field advantage of individual teams and that of the whole league as
\begin{eqnarray}\label{main_model_alpha}
 \beta_{i} = \Delta + u_i,\quad u_i\sim \mathcal{N}(0,\sigma^2), 
\end{eqnarray}
where $\sigma^2$ describes the variation of home field advantages across different teams. Based on \eqref{main_model_beta}, oppositely, we have the model for the match between team $T_j$ (with home field advantage) and team $T_i$ (without home field advantage) as
\begin{equation}\label{main_model_op}
    Y_{j,i}=-\alpha_{i,j}+  \beta_{j}+\epsilon_{j,i},
\end{equation}
where the net score without home field advantage satisfies $\alpha_{i,j} = - \alpha_{j,i}$. Thus, combining \eqref{main_model_beta} and \eqref{main_model_op}, we have \begin{equation}\label{hte}
 Y_{i,j} + Y_{j,i}= \beta_{i}+\beta_{j}+\epsilon_{i,j}+\epsilon_{j,i}.
\end{equation}
In other words, $\alpha_{i,j}$'s are treated as nuisance parameters and do not need to be estimated in our model. By repeating \eqref{hte} over all paired matches between team $T_i$ and team $T_j$, 
for $i\not = j$ and $i,j \in \{1,\cdots,n\}$, we have 
\begin{equation}\label{est3}
\underbrace{\begin{bmatrix}
1 & 1 & 0&\cdots &0&0&0\\
1 & 0 & 1&\cdots &0&0&0\\
\vdots & \vdots  & \vdots &\cdots &\vdots &\vdots &\vdots \\
0 & 0 & 0&\cdots &1&0&1\\
0 & 0 & 0&\cdots &0&1&1\\
\end{bmatrix}}_{\bm{H}_{N/2\times n}}
\underbrace{\begin{pmatrix}
  {\beta}_1\\
  {\beta}_2\\
 {\beta}_3\\
\vdots\\
 {\beta}_{n-2}\\
 {\beta}_{n-1}\\
 {\beta}_n
\end{pmatrix}}_{\widehat{\bm{\beta}}_{n\times 1}}
+ \underbrace{\begin{pmatrix}
\epsilon_{1,2} + \epsilon_{2,1}\\
\epsilon_{1,3} + \epsilon_{3,1}\\ 
\vdots\\
\epsilon_{(n-2),n} + \epsilon_{n,(n-2)}\\ 
\epsilon_{(n-1),n} + \epsilon_{n,(n-1)}
\end{pmatrix}}_{\bm{\epsilon}_{N/2\times 1}}
=
\underbrace{\begin{pmatrix}
Y_{1,2} + Y_{2,1}\\
Y_{1,3} + Y_{3,1}\\ 
\vdots\\
Y_{(n-2),n} + Y_{n,(n-2)}\\ 
Y_{(n-1),n} + Y_{n,(n-1)}
\end{pmatrix}}_{\bm{Y}_{N/2\times 1}}.
\end{equation} 
This motivates us to estimate the team-specific home advantage effects through the above linear equation system. While collapsing paired match outcomes removes the unobserved baseline term \(\alpha_{i,j}\) and improves identifiability of home field advantage (HFA), it also reduces the sample size from \(N\) to \(N/2\) and limits our ability to assess absolute team performance on a neutral field. As a result, we cannot disentangle skill-based differences from venue effects, and the method focuses solely on estimating relative home field advantage rather than intrinsic team strength.
Specifically, since $\bm{H}$ is a full rank matrix under $n\geq 3$, we use the following estimate of $\bm{\beta}=[\beta_1, \cdots,\beta_n]^\top$,
\begin{equation}
\widehat{\bm{\beta}} = (\bm{H}^\top \bm{H})^{-1} \bm{H} \bm{Y}.
\end{equation} 
Based on the definition that $\Delta=\sum_i \beta_i /n$, we have an estimator of  $\Delta$ as
\begin{equation}\label{est1}
    \widehat{\Delta} =  \sum_i \widehat{\beta}_i /n,  
\end{equation}
where $\widehat{\beta}_i$ is the $i$-th element in $\widehat{\bm{\beta}} $. Following the standard theory for linear regression, we establish the normality of $\widehat{\bm{\beta}} $ in the following proposition.

\begin{prop}
\label{beta_variance}
Assuming the noise terms $\epsilon_{i,j}$ are independent and identically distributed (i.i.d.) Gaussian random variables, that is, $\epsilon_{i,j}\sim \mathcal{N} (0,\sigma_0^2)$. Under (A1)-(A3) with $n\geq3$, we have   
\begin{equation*}
      \widehat{\bm{\beta}} - \bm{\beta}   \sim \mathcal{N}_{n}(\bm{0},\Sigma_{\bm{\beta}}), 
\end{equation*} 
where $\bm{\beta}=[\beta_1,\cdots,\beta_n]^\top$ denotes the true causal effect for $n$ teams. The covariance matrix $\Sigma_{\bm{\beta}}$ can be estimated by $\widehat{\sigma}^2_{\beta}(\bm{H}^{\top}\bm{H})^{-1}$, where $\widehat{\sigma}^2_{\beta}=2||\bm{H}\widehat{\bm{\beta}}-\bm{Y}||^2_{2}/N$.
\end{prop}
Proposition \ref{beta_variance} holds for every $n \geq 3$ since the normality of $\widehat{\beta}$ is exact. A two-sided $(1-\alpha)$ marginal confidence band  of $\bm{\beta}$ can be obtained as
\begin{equation}
    \widehat{\bm{\beta}}\pm z_{\alpha/2}\sqrt{\text{diag}\{\widehat{\sigma}^2_{\beta}(\bm{H}^{\top}\bm{H})^{-1})\}},
\end{equation}
where $z_{\alpha/2}$ denotes the upper $\alpha/2-$th quantile of a standard normal distribution,
using the estimation of $\Sigma_{\bm{\beta}}$ from Proposition \ref{beta_variance}. Since $\widehat{\Delta}$ is the sample mean of $\widehat{\beta}_i$ as indicated in \eqref{est1}, the normality of the estimated average treatment effect can be obtained immediately from Proposition \ref{beta_variance}.
\begin{prop}
\label{prop:sigma_delta}
Suppose that the same set of assumptions in Proposition \ref{beta_variance} hold. Let $\Delta = \sum_i \beta_i/n$. Then    
\begin{equation*}
    \sqrt {n}\{\widehat{\Delta} -\Delta\}  \sim  \mathcal{N}(0,\sigma^2),
\end{equation*}
where an unbiased estimator for the variance $\sigma^2$ is \begin{equation}\label{eq:est} 
\widehat{\sigma}^2 ={\frac{1}{n -1}} \sum_{i=1}^n (\widehat{\beta}_i - \widehat{\Delta})^2 -  \sum_{i=1}^n \text{diag}\{\widehat{\sigma}^2_{\beta}(\bm{H}^{\top}\bm{H})^{-1})\}_i /n,
\end{equation}
and $\text{diag}(W)_i$ is the $i$-th diagonal element of matrix $W$. 
\end{prop}
The first part of Proposition \ref{prop:sigma_delta} is obvious since $\widehat{\Delta}$ is a linear combination of $\widehat{\bm{\beta}}$, which follows a multivariate normal distribution. However, the estimation of $\sigma^2$ is non-trivial because the off-diagonal entries in the covariance estimator for $\Sigma_{\beta}$ do not perform well due to the high-dimensionality, i.e., the number of the free parameters in $\Sigma_{\beta}$ 
is $O(n^2)$ and our data have a sample size of the same order. Therefore, the usual quadratic form estimator for $\sigma^2$, $n^{-1} \widehat{\sigma}^2_{\beta} \bm{1}^{\top} (\bm{H}^{\top}\bm{H})^{-1})\bm{1}$, is biased, where $\bm{1}$ is the vector of $n$ ones. To solve this problem, the key observation is to use the law of total variance as suggested in the hierarchical modeling literature \citep{berry2013bayesian,chu2018bayesian,geng2020mixture} by considering
\begin{equation*}
 \sigma^2=\text{Var}\{\Mean_i(\widehat{\beta_i}|\bm{O})\}= \text{Var}_i(\widehat{\beta_i})-\Mean_i\{\text{Var}(\widehat{\beta_i}|\bm{O})\},
\end{equation*} 
where $\bm{O}$ is the observed data. The first term $\text{Var}_i(\widehat{\beta_i})$ can be estimated by the sample variance $  \sum_i(\widehat{\beta}_i - \widehat{\Delta})^2/(n -1)$, and the second term can be estimated by $\sum_i\text{diag}\{\widehat{\sigma}^2_{\beta}(\bm{H}^{\top}\bm{H})^{-1})\}_i /n$. Both estimators are unbiased. Hence \eqref{eq:est} holds. 
The two-sided $(1-\alpha)$ confidence interval of $\Delta$ thus can be constructed as 
\begin{equation}
    \widehat{\Delta}\pm z_{\alpha/2} \sqrt{\widehat{\sigma}^2/n},
\end{equation}
based on Proposition \ref{prop:sigma_delta}.
\begin{remark}
The normality of $\beta_i$ in \eqref{main_model_alpha} can be relaxed to other distributions with mean $\Delta$ and variance $\sigma^2$, which leads to an asymptomatic normality of $\widehat{\Delta}$ as $\sqrt {n}\{\widehat{\Delta} -\Delta\}  \stackrel{d}{\rightarrow} \mathcal{N}(0,\sigma^2)$ when $n\to \infty$, by the central limit theorem. Yet, in reality, we have a finite and usually fairly small number of teams in one league, such as $n=20$ for EPL. Therefore we choose to keep the normality assumption. 
\end{remark}
\subsection{Multivariate Regression Extension}

Consider $K$ summary statistics together, denote $\mB=(\bm{\beta}^{(1)},\ldots,\bm{\beta}^{(K)})$, $\mY=(\bm{Y}^{(1)},\ldots,\bm{Y}^{(K)})$, and $\mE=(\bm{\epsilon}^{(1)},\ldots,\bm{\epsilon}^{(K)})$. The multivariate linear regression can be written as
\begin{equation}
    \mY=\bm{H}\mB+\mE,
\end{equation}
where $\mY$ is $N/2\times K$ response matrix, $\mB$ is $n\times K$ matrix of coefficients, and $\mE$ is $N/2\times K$ error matrix. We use the following estimate of $\mB$
\begin{equation}
\widehat{\mB} = (\bm{H}^\top \bm{H})^{-1} \bm{H}^\top \mY.
\end{equation} 
Based on the hierarchical structure of the home field advantage of individual teams and that of the whole league, the league level home field advantage of $K$ summary statistics, $\mathcal{D}=(\Delta_1,\ldots,\Delta_K)^\top$ can be estimated by  
\begin{equation}
    \widehat{\mathcal{D}}=\text{colSums}(\widehat{\mB})/n,
\end{equation}
where $\text{colSums}(\bm{A})$ is finding sum of each column of the matrix $\bm{A}$. Following the standard theory for multivariate linear regression, we
establish the normality of $\widehat{\mB}$ in the following proposition.

\begin{prop}
\label{B_variance}
Assuming the noise terms $\bm{\epsilon}^{(j)}, j=1,\ldots,K$ follows a multivariate Gaussian distribution as
\begin{equation*}
   \bm{\epsilon}^{(j)}\sim  \mathcal{N}_{N/2}(\bm{0},\sigma^2_j\bm{I}),
\end{equation*}
where $\bm{\epsilon}^{(j)}$ is the $j$-th column of $\mE$, and $\bm{\epsilon}_j^\top, j=1,\ldots,N/2$ follows a
multivariate Gaussian distribution as
\begin{equation*}
   \bm{\epsilon}^\top_j\sim  \mathcal{N}_{K}(\bm{0},\Sigma_e),
\end{equation*}
where  $\bm{\epsilon}_j$ is the $j$-th row of  $\mE$. Under (A1)-(A3) with $n\geq3$, we have 
\begin{equation*}
  vec(\widehat{\mB}^\top-\mB^\top)\sim \mathcal{N}_{nK}(\bm{0}, \Sigma_{\mB}),
\end{equation*}
where $\mB$ denotes the true causal effect for $n$ teams with $K$ summary statistics and $vec(\mB^\top)=(\bm{\beta}_1,\ldots,\bm{\beta}_n)^\top$ and $\bm{\beta}_j,j=1,\ldots n$ is the $j$-th row of $\mB$. The error covariance matrix $\Sigma_{\mB}$ can be estimated by $\widehat{\Sigma}_{\mB}=(\bm{H}^\top\bm{H})^{-1}\otimes\frac{\mY^\top(\bm{I}_{N/2}-\bm{H}(\bm{H}^\top \bm{H})^{-1} \bm{H}^\top)\mY}{N/2-n}$.
\end{prop}
A two-sided $(1-\alpha)$ marginal confidence band  of $vec(\mB^\top)$ can be obtained as
\begin{equation}
    vec(\widehat{\mB}^\top)\pm z_{\alpha/2}\sqrt{\text{diag}\{{\Sigma}_{\mB}\}},
\end{equation}
where $z_{\alpha/2}$ denotes the upper $\alpha/2-$th quantile of a standard normal distribution,
using the estimation of ${\Sigma}_{\mB}$ from Proposition \ref{B_variance}.

\begin{prop}
\label{prop:sigma_D}
Suppose that the same set of assumptions in Proposition \ref{B_variance} hold. Then    
\begin{equation*}
    \sqrt {n}\{\widehat{\mathcal{D}} -\mathcal{D}\}  \sim  \mathcal{N}_K(0,\Sigma_{\mathcal{D}}),
\end{equation*}
where $\mathcal{D}$ denotes the true league level causal effects for $K$ summary statistics and an unbiased estimator for the covariance matrix $\Sigma_{\bm{D}}$ is 
\begin{equation}\label{eq:est_D}
\widehat{\Sigma}_{\bm{D}}={\frac{1}{n - K}} \sum_{i=1}^n (\widehat{\bm{\beta}}_i - \widehat{\mathcal{D}}^\top)^\top  (\widehat{\bm{\beta}}_i - \widehat{\mathcal{D}}^\top) -  \sum_{i=1}^n \text{blockdiag}\{\widehat{\Sigma}_{\mB}\}_i /n,
\end{equation}
and $\text{blockdiag}(W)_i$ is the $i$-th $K\times K$ diagonal matrix of matrix $W$. 
\end{prop}
The two-sided $(1-\alpha)$ confidence interval of $\mathcal{D}$ thus can be constructed as 
\begin{equation}
    \widehat{\mathcal{D}}\pm z_{\alpha/2} \sqrt{\text{diag}\{\widehat{\Sigma}_{\bm{D}}\}/n},
\end{equation}
based on Proposition \ref{prop:sigma_D}.

\section{Simulation}\label{sec:simu}
\subsection{Simulation Setup: Univariate response case}
In this section, simulation studies are carried out to evaluate the empirical performance of estimators for both $\bm{\beta}$'s and $\Delta$. We consider two data generation scenarios for $\alpha_{ij}$ in our simulation. In the first scenario, we generate the ability difference between team $T_i$ and team $T_j$ as $\alpha_{ij}\overset{\text{i.i.d}}{\sim} \mathcal{N}(0,2^2), i=1,\ldots,n,j=1,\ldots,n$ and $\alpha_{ij}=-\alpha_{ji}$. In second scenario, we first generate the ability of each team  $\text{Ab}_{i}\sim \mathcal{N}(0,2^2),i=1,\ldots,n$, and then set $\alpha_{ij}=\text{Ab}_{i}-\text{Ab}_{j}$ for $i=1,\ldots,n,\, j=1,\ldots,n$. We fix $\Delta=1$ for all the simulation studies, $\beta_i\overset{\text{i.i.d}}{\sim} \mathcal{N}(1,0.3^2)$, and $\epsilon_{ij}\overset{\text{i.i.d}}{\sim} \mathcal{N}(0,\sigma^2_0), i=1,\ldots,n,\, j=1,\ldots,n$. We choose three different values for the variance of $\epsilon_{ij}$, $\sigma^2_0=0.5,1,2$, and four different values for the number of teams, $n=10,20,40,80$. In total, we generate $1,000$ independent replicates for each scenario. The performance of the estimates are evaluated by the bias, the coverage probability (CP), the sample variance of the 1,000 estimates (SV), and the mean of the 1,000 variance estimates (MV) in the following way; take $\Delta$ as an example:
\begin{equation*}
\begin{split}
&\text{Bias} = \frac{1}{1000} \sum_{i=1}^{1000}
\widehat{\Delta}^{(i)} - \Delta^{(0)} , \\
&\text{CP} = 
\frac{1}{1000}1\left\{\Delta^{(0)}\in (\Delta^{(i)}_{\alpha/2},\Delta^{(i)}_{1-\alpha/2})\right\},\\
&\text{SV} =  \frac{1}{1000-1}
\sum_{i=1}^{1000}
 \left(\widehat{\Delta}^{(i)} -\bar{\widehat{\Delta}} \right)^2,\\
 &\text{MV} =  \frac{1}{1000}
\sum_{i=1}^{1000}
\widehat{\text{Var}}(\Delta^{(i)}) ,
   \end{split} 
\end{equation*}
where $\Delta^{(0)}$ is the true value for $\Delta$, $\widehat{\Delta}^{(i)}$, $ (\Delta^{(i)}_{\alpha/2},\Delta^{(i)}_{1-\alpha/2})$, and $\widehat{\text{Var}}(\Delta^{(i)})$ are the point estimate, $(1-\alpha)$-level confidence bands, and variance estimate from the $i$th replicated simulation data, respectively. We set $\alpha = 0.05$ throughout the simulation.

\subsection{Simulation Results: Univariate response case}
We first examine the simulation results for $\Delta$, as well as the variance estimator of $\Delta$. In the supplement, Table~\ref{tab:delta_results} summarizes the performance of the $\Delta$ estimator in different data generation scenarios, and reports the bias, coverage probability, sample variance
of the 1,000 estimates and mean of the 1,000 variance estimates. From this table, it is
clear that our proposed estimator for $\Delta$ has a very small bias that decreases quickly as $n$ increases. The coverage probabilities are also very close to the desired $95\%$ nominal coverage level, especially when $n \geq 20$. Note that the standard error of the estimated coverage probabilities is $\sqrt{.05\times.95/1,000}=0.0069$. The last two columns also confirm the accurate variance estimation for $\widehat{\Delta}$ under all the simulation scenarios; and the estimation accuracy becomes better as $n$ increases.  

\begin{figure}[!h]
    \centering
    \includegraphics[width=0.95\textwidth]{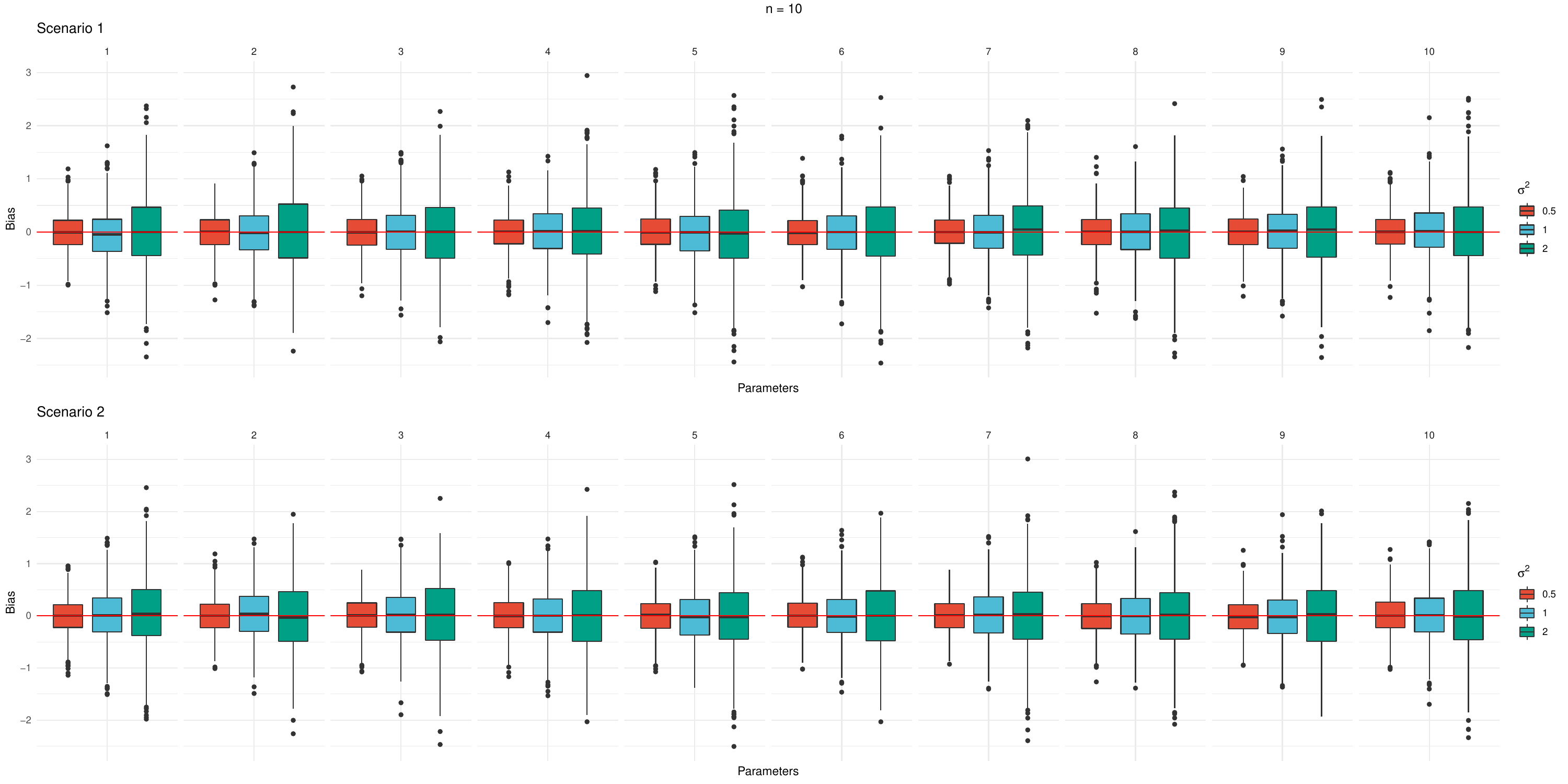}

    \caption{Simulation results: boxplot for Bias of $\widehat{\bm{\beta}}$ when $n=10$.}
    \label{fig:bias_beta1}
\end{figure}

\begin{figure}[!h]
    \centering
    \includegraphics[width=0.95\textwidth]{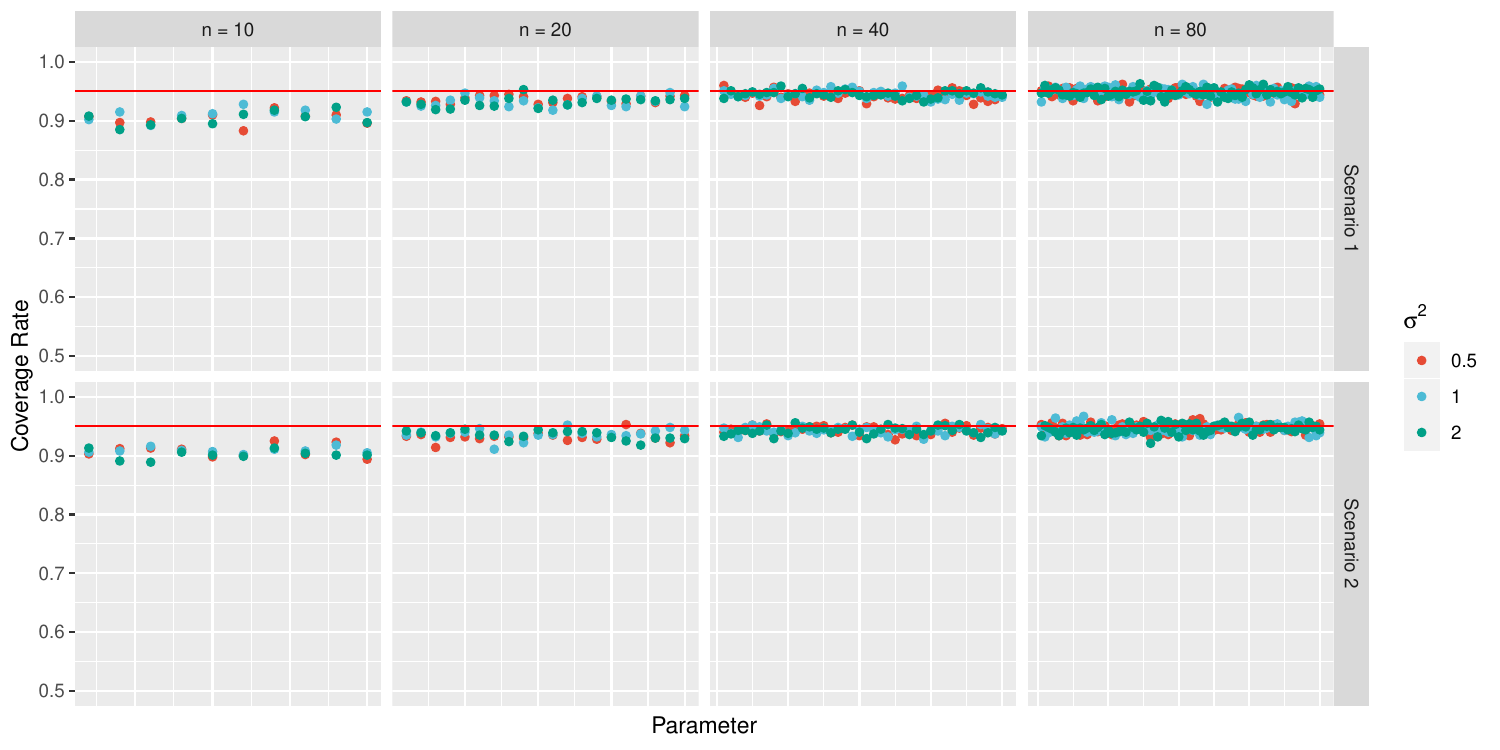}
    \caption{Simulation: Coverage Probabilities for $\bm{\beta}$'s in different scenarios.}
    \label{fig:cover_beta}
\end{figure}

Next we examine the simulation results for $\bm{\beta}$'s, as well as the
coverage probability of $\bm{\beta}$'s. Figure \ref{fig:bias_beta1} summarizes the boxplots of all estimators $\bm{\beta}$ in different data generation processes to show the estimation performance. It is
clear from the figure that the estimator is essentially unbiased for all scenarios considered in the simulation. Furthermore, we present the empirical coverage probabilities for the confidence intervals in Figure~\ref{fig:cover_beta}. All the coverage probabilities are fairly close to the nominal level of 0.95, especially when $n \geq 20$. The coverage probability in general becomes more accurate  
as the sample size $n$ increases.

In summary, the simulation results confirm the excellent performance (e.g., bias, variance estimate, and coverage probabilities) of our method in terms of estimating both $\Delta$ and $\bm{\beta}$'s. Note that our simulation scenarios also include the situation when $\alpha_{ij}$ are not $i.i.d.$ generated. Our estimator still performs well in this case because our estimation procedure does not require modeling assumptions or estimation of  $\alpha_{ij}$. The simulation results validate the theoretical results, especially the variance formula, in Propositions \ref{beta_variance} and \ref{prop:sigma_delta}.

\subsection{Simulation Setup: Multivariate Response Case}
In this section, simulation studies are conducted in the multivariate response case. We fix  $\mathcal{D}=\mathbf{1}_K$ and $K=3$ for all the simulation studies, $\mathcal{B}_i\overset{\text{i.i.d}}{\sim} \mathcal{N}(\mathcal{D},\Sigma_b)$, and $\mathcal{E}_i\overset{\text{i.i.d}}{\sim} \mathcal{N}(\mathbf{0}_K,\Sigma_E), i=1,\ldots,N/2$, where $\Sigma_b$ and $\Sigma_E$ follow an AR structure:
\[
\Sigma_{ij} = \rho^{|i-j|}, \quad \rho = 0.3, \quad i,j = 1,\dots,K.
\] We choose three different values for the number of teams, $n=10,20,40$. In total, we implement $1,000$ independent replicates for each scenario.

\subsection{Simulation Results: Multivariate Response Case}
Table \ref{tab:delta_results_multivar} summarizes the performance of the estimator $\mathcal{D}$, and reports the bias, mean bias in all 3 dimensions, coverage probability, and mean probability across all 3 dimensions. In multivariate response cases, it clearly states that our proposed estimator for $\mathcal{D}$ has a very small bias, the overall mean of which decreases in magnitude with increasing $n$. The coverage probabilities are also close to the nominal level of 0.95, especially when $n \geq 20$. From the table, the mean coverage probabilities are closer to the nominal coverage level 95\% as $n$ increases.

\begin{table}[h]
\centering
\caption{Summary of simulation results for $\widehat{\mathcal{D}}$. (Column (1): bias over 1000 simulations, (2): mean bias across all 3 dimensions, (3): coverage probabilities over 1000 estimates, (4): mean coverage probabilities across all dimensions).}\label{tab:delta_results_multivar}
\begin{tabular}{ccccc}
\toprule
 & (1) Bias & (2) Mean Bias & (3) CP & (4) Mean CP \\ 
  \midrule
n=10 & (0.00136, -0.00887, -0.01262) & -0.00671 & (0.929, 0.924, 0.927) & 0.927 \\ 
  \midrule
  n=20 & (0.00382, 0.00241, 0.00790) & 0.00471 & (0.945, 0.944, 0.948) & 0.946 \\ 
  \midrule
  n=40 & (-0.00043, 0.00583, 0.00594) & 0.00378 & (0.943, 0.951, 0.953) & 0.949 \\ 
\bottomrule
\end{tabular}
\end{table}

Finally, we examine the simulation results of $\widehat{\mathcal{B}}$ in multivariate response cases. We pick the first 3 teams in Table \ref{tab:beta_results_multivar_bias} and \ref{tab:beta_results_multivar_cp}, showing the bias, coverage probability, and their means in all dimensions. Our estimator $\widehat{\mathcal{B}}$ shows a very small bias. Furthermore, all coverage probabilities are close to the desired nominal level 95\%.

\begin{table}[ht]
\centering
\caption{Summary of simulation results of $\widehat{\mathcal{B}}$ for the first 3 teams. Column (1): bias over 1000 simulations; (2): mean bias across all 3 dimensions.}\label{tab:beta_results_multivar_bias}
\begin{tabular}{clcc}
\toprule
 & & (1) Bias & (2) Mean Bias \\ 
  \midrule
n=10 & Team 1 & (-0.012851, -0.010288, 0.017544) & -0.001865 \\ 
    & Team 2 & (0.007158, -0.003525, -0.010501) & -0.002289 \\ 
    & Team 3 & (-0.007598, 0.000614, 0.001208) & -0.001925 \\ 
  \midrule
  n=20 & Team 1 & (-0.007175, -0.012686, 0.004830) & -0.005011 \\ 
    & Team 2 & (-0.002764, 0.014926, -0.006162) & 0.002000 \\ 
    & Team 3 & (0.016416, 0.006142, 0.002108) & 0.008222 \\ 
  \midrule
  n=40 & Team 1 & (-0.001944, 0.004321, -0.003710) & -0.000445 \\ 
    & Team 2 & (-0.001809, 0.000182, 0.001515) & -0.000037 \\ 
    & Team 3 & (-0.011155, -0.009673, -0.006627) & -0.009152 \\ 
\bottomrule
\end{tabular}
\end{table}

\begin{table}[ht]
\centering
\caption{Summary of simulation results of $\widehat{\mathcal{B}}$ for the first 3 teams. Column (1): coverage probabilities over 1000 estimates; (2): mean coverage probabilities across all dimensions.}\label{tab:beta_results_multivar_cp}
\begin{tabular}{clcc}
\toprule
 & & (1) CP & (2) Mean CP \\ 
  \midrule
n=10 & Team 1 & (0.945, 0.936, 0.939) & 0.940 \\ 
    & Team 2 & (0.943, 0.956, 0.951) & 0.950 \\ 
    & Team 3 & (0.947, 0.943, 0.938) & 0.943 \\ 
  \midrule
  n=20 & Team 1 & (0.945, 0.948, 0.951) & 0.948 \\ 
    & Team 2 & (0.934, 0.937, 0.949) & 0.940 \\ 
    & Team 3 & (0.947, 0.964, 0.946) & 0.952 \\ 
  \midrule
  n=40 & Team 1 & (0.953, 0.945, 0.942) & 0.947 \\ 
    & Team 2 & (0.955, 0.944, 0.953) & 0.951 \\ 
    & Team 3 & (0.945, 0.952, 0.955) & 0.951 \\ 
\bottomrule
\end{tabular}
\end{table}

In summary, the simulation results demonstrate that our proposed estimation methods perform exceptionally well in multivariate response settings. Specifically, the estimators exhibit low bias and achieve nominal coverage probabilities across a range of scenarios, confirming the reliability and robustness of our approach under realistic data-generating conditions.

\section{Analysis of EPL Data}\label{sec:app}
\subsection{Independent Analysis of EPL Data}
\begin{figure}[!t]
    \centering
    \includegraphics[width=1\textwidth]{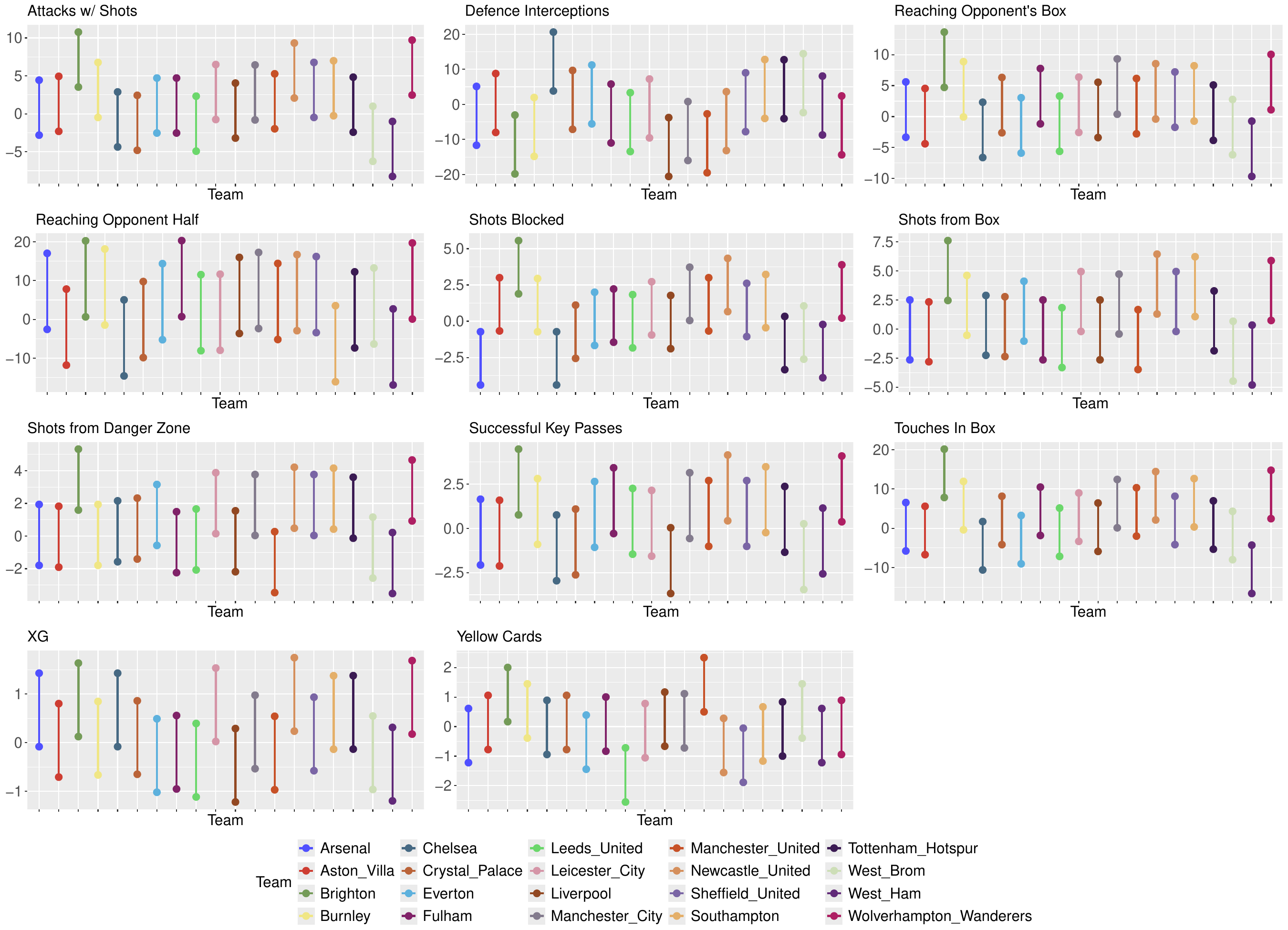}
    \caption{EPL data analysis: estimated $\bm{\beta}$ with confidence intervals for all in-game statistics.}
    \label{fig:cover_beta1}
\end{figure}

We focus on the 11 selected in-game statistics collected from 20 teams in the English Premier League in 2020-2021 as described in Section \ref{sec:data}. For each statistic, we treat it as the main outcome and apply the proposed causal inference method to calculate the team-specific home field advantage $\widehat{\beta}_i$ and the league-wide home field advantage $\widehat{\Delta}$  to further analyze the implications of the home field advantage on a team-by-team basis and a statistic-specific basis. 
Figure~\ref{fig:cover_beta1} shows the 95\% confidence intervals for the $\widehat{\beta}_i$. A $\beta=0$ would indicate no home field advantage. Brighton is clearly a standout team here, with a wide confidence interval across all statistics. 

\begin{figure}[!t]
    \centering
    \includegraphics[width=1.1\textwidth]{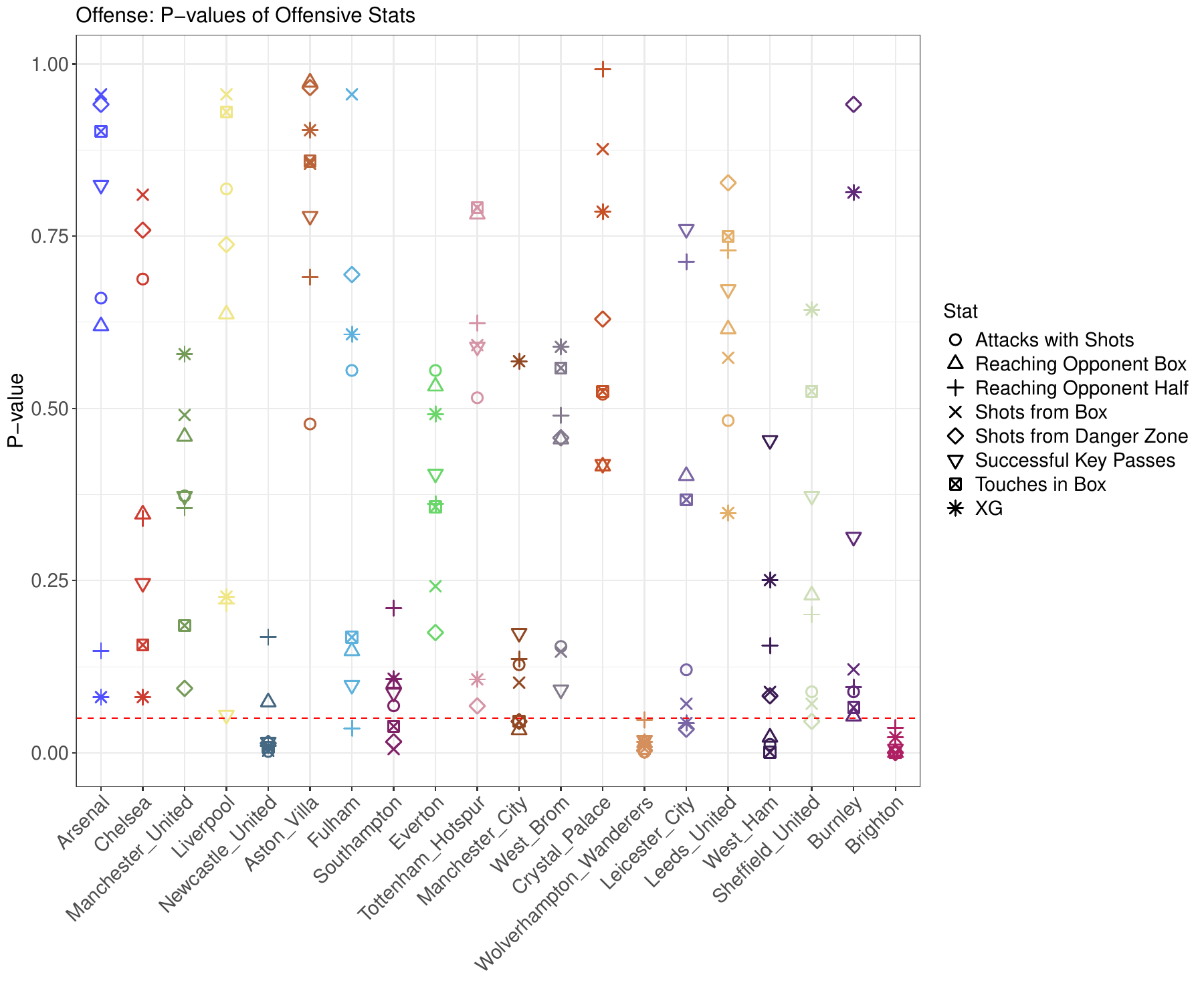}
    \caption{EPL data analysis: p-values for $\bm{\beta}$ estimators of 8 offensive statistics.}
    \label{fig:pvalue_beta_off}
\end{figure}

\begin{figure}[!t]
    \centering
    \includegraphics[width=1.1\textwidth]{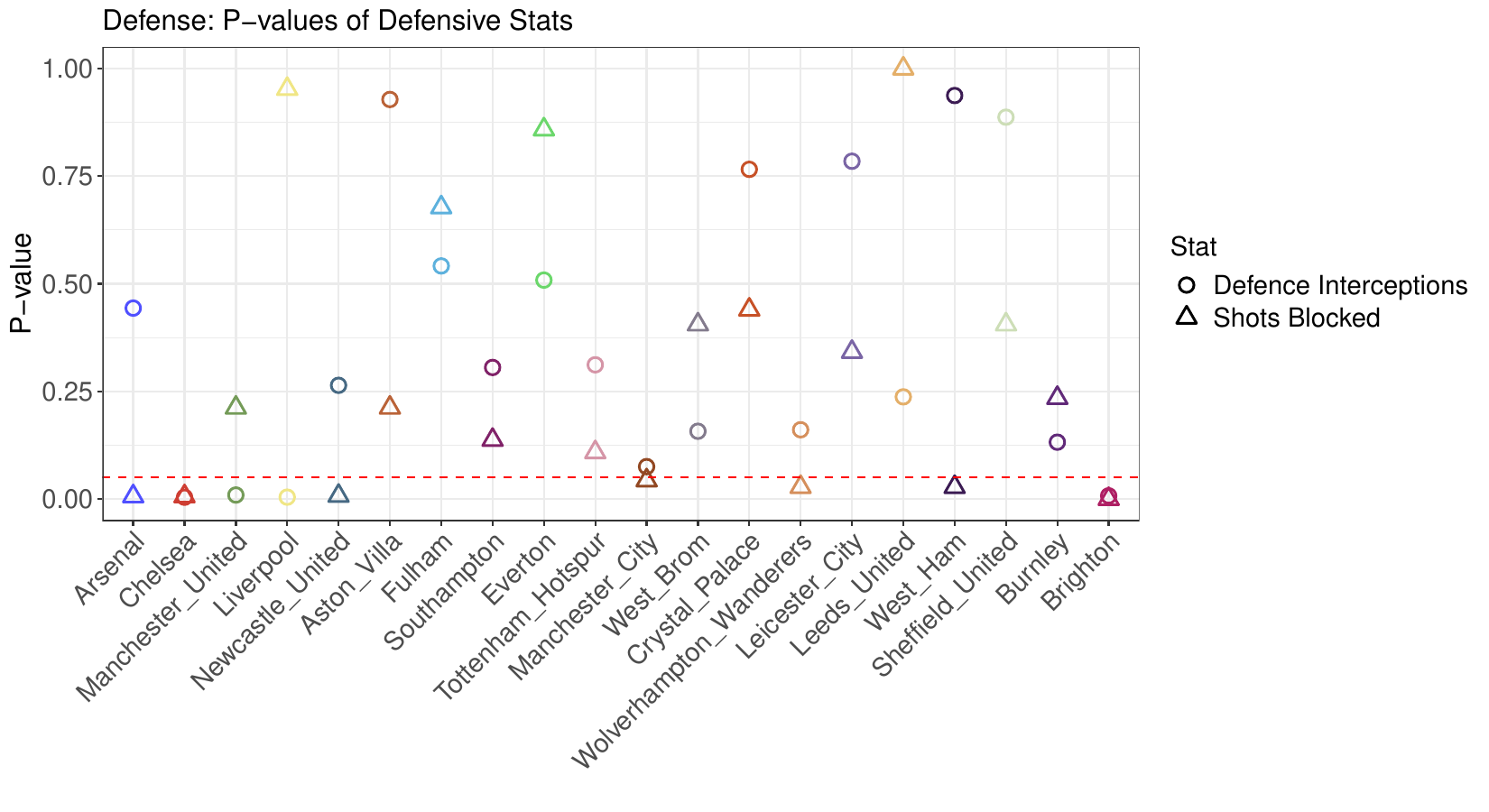}
    \caption{EPL data analysis: p-values for $\bm{\beta}$ estimators of 2 defensive statistics.}
    \label{fig:pvalue_beta_def}
\end{figure}

\begin{figure}[!t]
    \centering
    \includegraphics[width=1.1\textwidth]{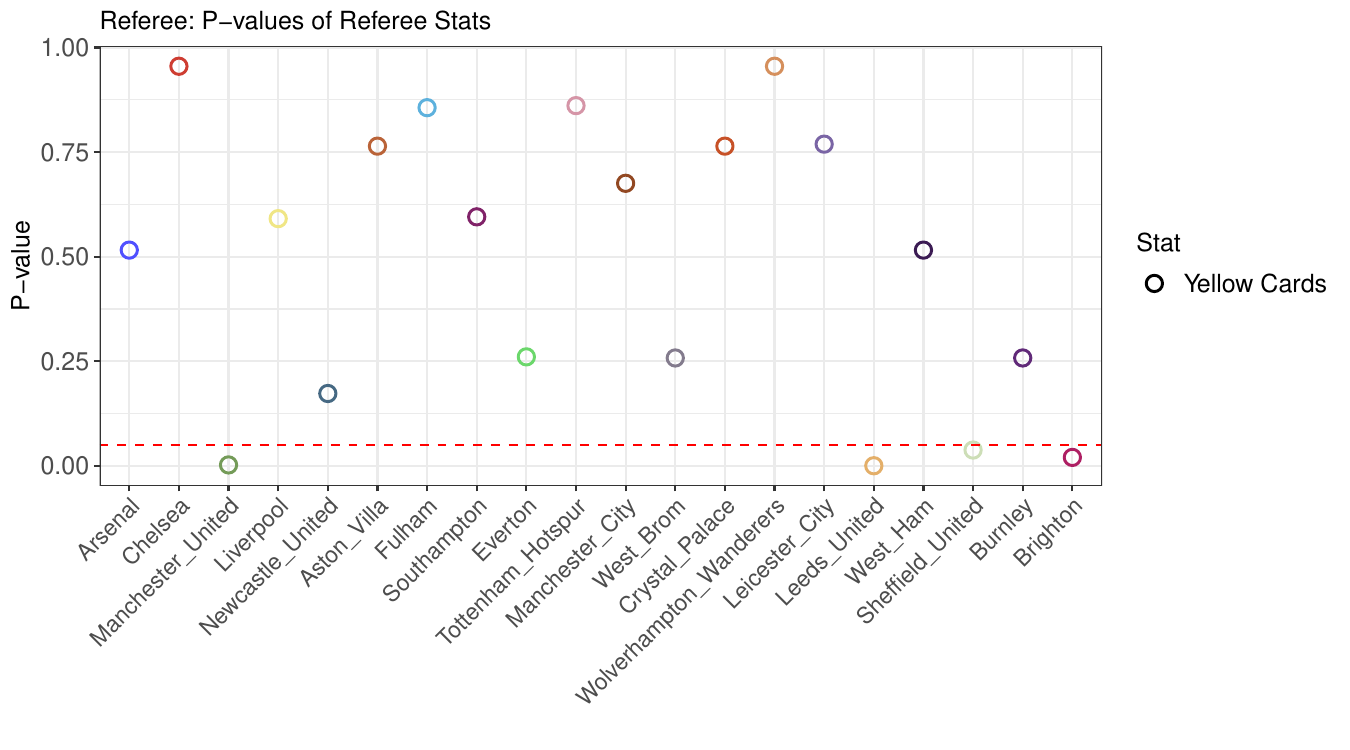}
    \caption{EPL data analysis: p-values for $\bm{\beta}$ estimators of the referee statistic.}
    \label{fig:pvalue_beta_ref}
\end{figure}

In Figure~\ref{fig:cover_beta1}, each of the eleven statistics had a range of 3-6 teams out of 20 that were significant for that particular statistic. While this is clearly not a majority, it does give us some data on the type of statistic that retains the home field advantage, which we will discuss more in the next section. Conversely, we can look at the significance of the $\widehat{\beta}_i$ by team, which will continue to build on to the story. The Figure~\ref{fig:pvalue_beta_off}, ~\ref{fig:pvalue_beta_def} and~\ref{fig:pvalue_beta_ref} shows the p-values of the $\widehat{\beta}_i$'s for each team (x-axis and color) and for each in-game statistic (denoted by shape), in 3 aspects of offense, defense, and referee respectively.

While a majority of the statistics are above the 0.05 standard threshold for significance, there are some teams that have a majority of the in-game statistics being significant. We now look at the makeup of those particular teams and begin to make inference on what makes those specific teams susceptible to the phenomenon that is the home field advantage. The teams that had the highest number of significant statistics were Fulham, Brighton, Newcastle United and Wolverhampton Wanderers. This prompted us to ask what these teams had in common that the other teams did not that caused them to reap the benefits of the home field advantage. The most obvious and telling common factor amongst the teams was their records- they all were ranked in the bottom $50\%$ at the end of the season, with Fulham being one of the three relegation teams in the English Premier League that year. This could imply that less performant teams reap higher benefits of home-field advantage than teams that excel. Teams that perform in the top rankings of the English Premier League can dominate their opponents regardless of their match location. Perhaps their talent and skill prevails such that the difference in their statistics when they are home versus when they are away is nearly indiscernible. In contrast, teams that are not as talented or skilled retain every advantage from being at home - from increased confidence due to crowd involvement, to familiarity, and to a lack of travel fatigue.

The $\widehat{\Delta}$ values tell us the estimated home field advantage that a specific statistic awards in the league. Significant $\widehat{\Delta}$ values give us insight to the statistics that have the highest impact on the home field advantage and how much they contribute. Table~\ref{tab:delta_estimate} shows the estimated $\Delta$ values for each of the in-game statistics, their estimated standard deviations, and their p-values for significance.

\begin{table}[!t]
    \centering
        \caption{EPL data analysis: estimated $\Delta$ for eleven summary statistics ordered by p-values (offensive statistics: black, defensive statistics: blue, referee statistics: red).}\label{tab:delta_estimate}
    \begin{tabular}{ c c c c c } 
  \toprule
  & \textbf{Statistic} & $\widehat{\Delta}$ & $\widehat{\sigma}$ & \textbf{P-value} \\
  \midrule
   Offense & Reaching Opponent Half & 3.592 & 0.949 & 0.000\\
  & Reaching Opponent Box & 1.732 &  0.608 & 0.004\\
  & Shots from Danger Zone & 0.786 & 0.279 & 0.005\\
   & Attacks w/ Shot & 1.568 & 0.556 & 0.005 \\
  & Shots from Box & 1.066 & 0.379 & 0.005 \\
  & XG & 0.232 & 0.092 & 0.011 \\
  & Touches in Box & 2.253 & 1.043 & 0.031 \\
  & Successful Key Passes &  0.489 & 0.237 & 0.039\\
  \midrule
  Defense & Defence Interceptions & -1.997 & 1.200 & 0.096  \\
  & Shots Blocked & 0.350 & 0.329 & 0.287\\
  \midrule
  Referee & Yellow Cards & -0.026 & 0.126 & 0.834 \\
  \bottomrule
\end{tabular}

\end{table}

Seven of the eleven statistics we chose to analyze proved to significantly have an effect on the net increase (or decrease) of a statistic in favor of the home team. 
We notice that offense based statistics, such as attacks with shot and reaching opponent box, are significant at $\alpha=0.05$; while defense based statistics, such as shots blocked and defence interceptions, are not significant, as well as the referee based statistic, yellow cards. This could be an indication that teams retain an advantage offensively when they compete at their home field. Going back to the causes of home-field advantage and examining the statistics, it appears that familiarity of the field would have an impact on the outcomes of the offensive statistics. For example, attacks with shot, reaching the opponent's half or box, shots from box or danger zone, and successful key passes are statistics that are based on the players' ability to get into scoring position, which would increase their likelihood of making a goal. The statistic goal itself did not show to hold a significant home field advantage, which allows us to infer that perhaps the number of goals scored by the home team may not be more than that of the away team, but the opportunities that are presented due to the causal factors of the home field advantage phenomenon are significantly greater. That is, the quality of play is better for the home team than the away team.

\subsection{Joint Analysis of EPL Data}
\begin{figure}[h]
    \centering
    \includegraphics[width=1\textwidth]{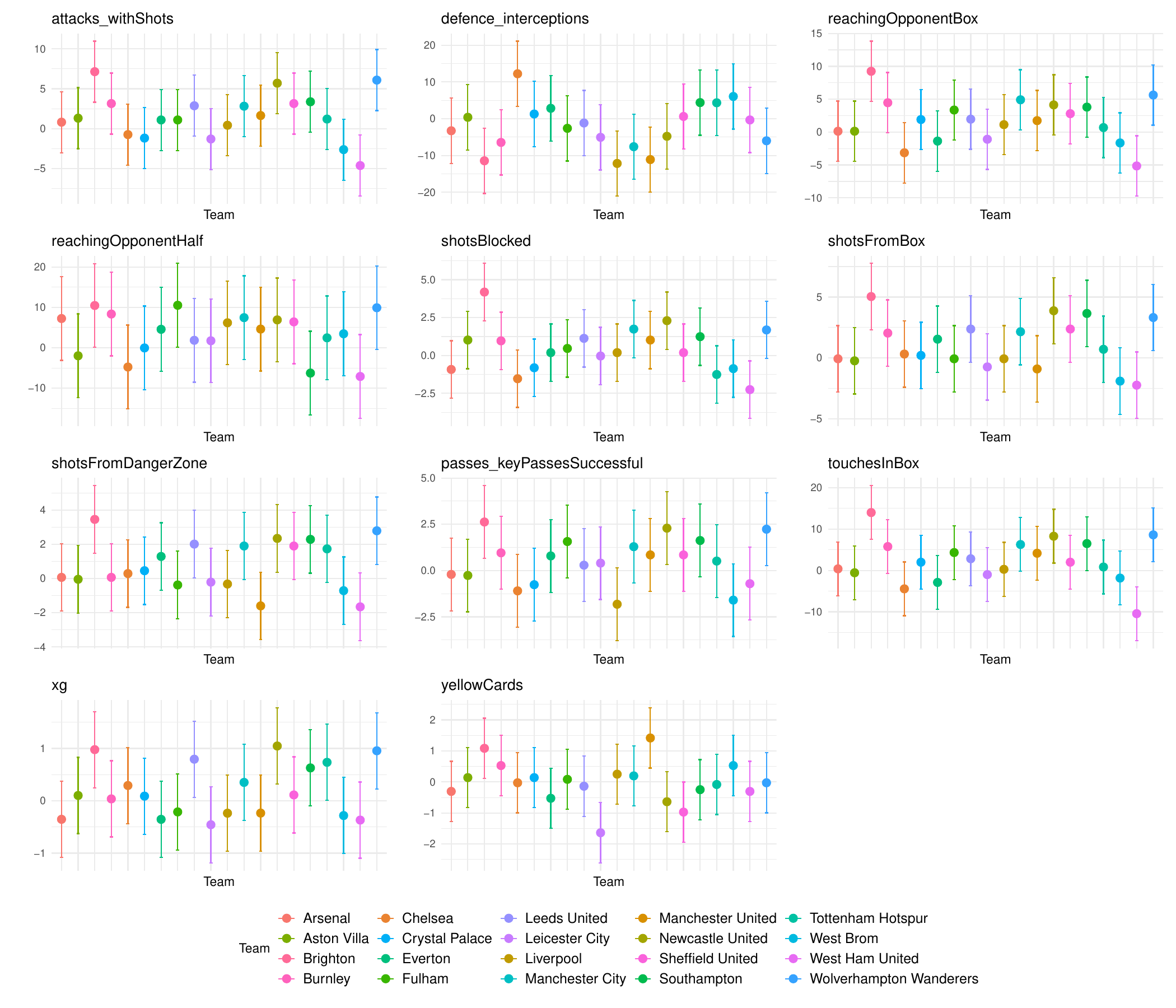}
    \caption{Estimated $\mB$ with confidence intervals of 20 teams grouped by in-game statistics: for each of the 11 performance metrics, we plot the estimated $\mB$ and 95\% confidence intervals across all 20 teams.}
    \label{fig:CI_B_Team_ByStat}
\end{figure}

We focus on multiple comparisons by jointly modeling in-game statistics collected from 2020-2021 English Premier League. Considering 11 summary statistics together as the multivariate outcome, we apply the proposed causal inference method for multivariate regression to compute the team-specific home field advantage $\widehat{\mathcal{B}}$ and the league-level home field advantage $\widehat{\mathcal{D}}$.
Figure~\ref{fig:CI_B_Team_ByStat} shows the 95\% confidence intervals for home field advantage estimates of 20 teams, grouped by performance metric.

\begin{figure}[!t]
    \centering
    \includegraphics[width=0.95\textwidth]{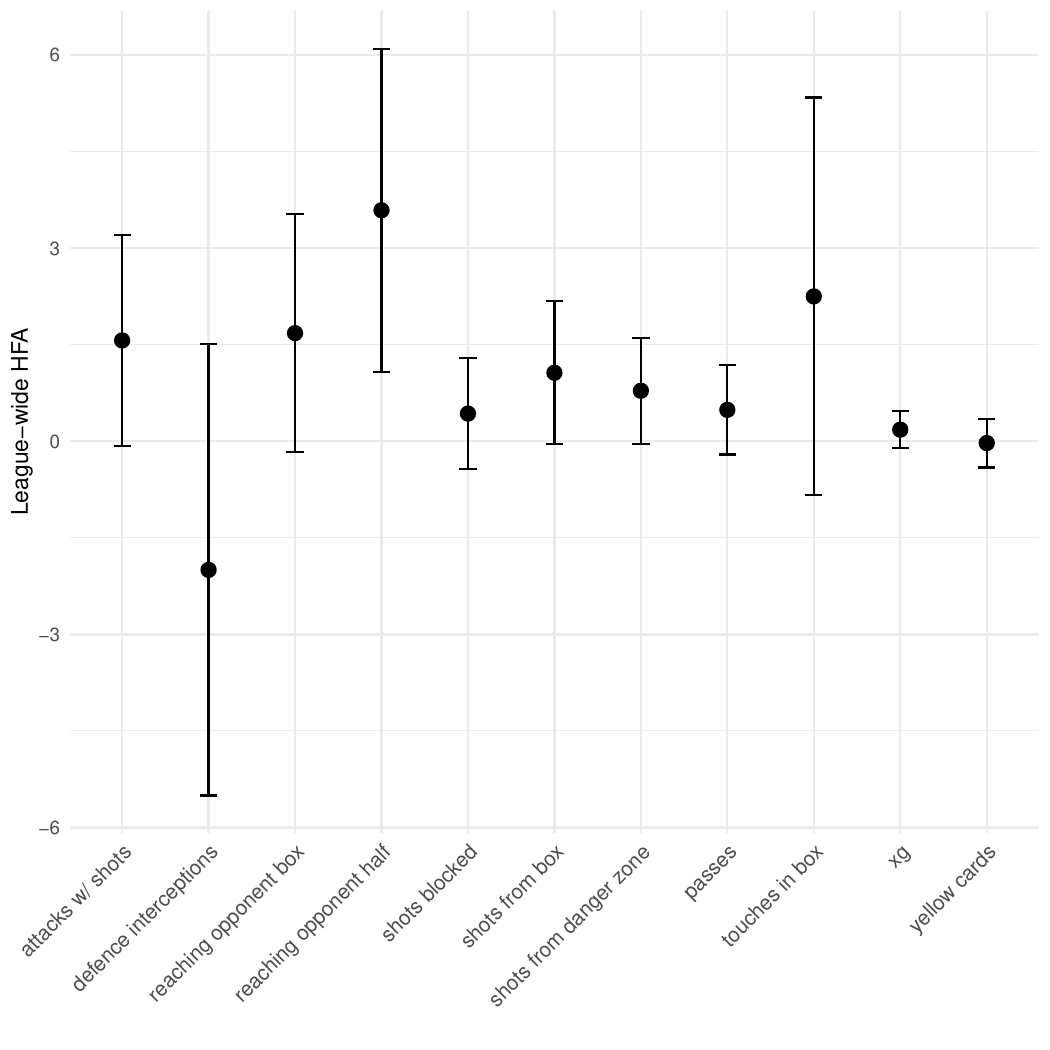}
    \caption{Estimated league-wide HFA with confidence intervals: estimated $\bm{D}$ and 95\% confidence intervals.}
    \label{fig:CI_D_LeagueStat}
\end{figure}

\begin{figure}[!t]
    \centering
    \includegraphics[width=0.95\textwidth]{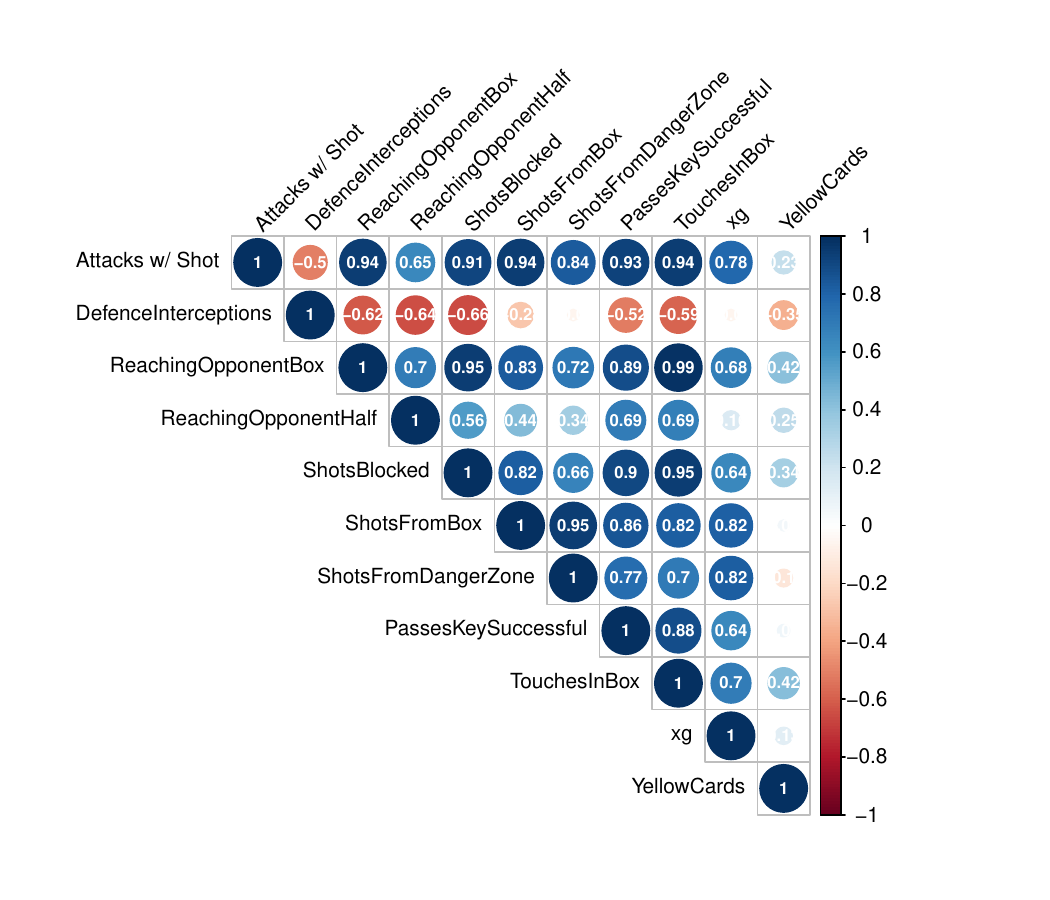}
    \caption{Correlation among 11 in-game statistics.}
    \label{fig:league_stat_cov}
\end{figure}

Figure~\ref{fig:CI_D_LeagueStat} shows the 95\% confidence intervals for league-wide home field advantages. We explicitly observe the correlations among the 11 metrics in Figure~\ref{fig:league_stat_cov}.

At the league level, home field advantage (HFA) shows strong and statistically significant effects in several offensive performance metrics. Notably, teams playing at home are more likely to reach the opponent’s half and penalty box, initiate attacks with shots, and generate attempts from dangerous areas. These findings suggest that the home setting facilitates forward progression and the creation of scoring opportunities. In contrast, metrics such as expected goals (xG) and yellow cards display minimal or statistically uncertain HFA effects, implying that shot quality and officiating outcomes are less influenced by game location. Defensive interceptions even show a slight negative HFA effect, potentially reflecting a strategic shift away from reactive play in favor of offensive engagement when playing at home.

Team-specific analyses further reveal substantial variation in HFA across clubs. For instance, Newcastle United (12th), Brighton (16th), and Sheffield United (20th) during the 2020--2021 season all exhibited pronounced home advantage effects in several performance dimensions. Meanwhile, clubs like Manchester City (1st), Manchester United (2nd), and Liverpool (3rd) showed smaller or negligible effects across most metrics, including expected goals and touches in the box. This pattern points toward a potential interaction between team-specific home field advantage parameters $\beta_j$ and baseline matchup expectations $\alpha_{ij}$. While the current model treats $\beta_j$ as independent of $\alpha_{ij}$, the empirical variation observed suggests that a more flexible specification---such as allowing $\beta_j$ to vary with expected neutral-field performance---may better capture underlying dynamics in team behavior. The team rankings for that season are provided in the supplementary materials.

Lastly, the multivariate joint modeling approach is well-supported by the observed correlation structure among match statistics. Offensive metrics such as attacks with shots, xG, touches in the box, and successful key passes are strongly correlated, indicating shared underlying dynamics. In contrast, defensive metrics and yellow cards are only weakly correlated with these offensive variables. These findings reinforce the value of modeling multiple outcomes jointly, which enhances estimation efficiency while allowing for nuanced interpretation of domain-specific effects. Overall, the results validate the multivariate framework and motivate further model extensions to account for team-level heterogeneity in home field advantage across different types of performance metrics.

\section{Discussion}\label{sec:disc}
Our study shows that home field advantage (HFA) in professional soccer is most evident in offensive performance metrics. Using a hierarchical causal model and multivariate joint analysis of eleven game-level statistics, we found that teams playing at home perform better in actions like successful key passes, attacks with shots, and touches in the opponent’s box. These subtle aspects of play reflect tactical and spatial advantages that may not appear in final scores but meaningfully shape match dynamics. In contrast, defensive metrics (e.g., interceptions) and referee-related outcomes (e.g., yellow cards) showed weak or insignificant HFA, suggesting that playing at home primarily boosts proactive offensive behavior rather than reactive or officiated aspects. Encouragingly, we found no evidence of referee bias across venues. At the team level, HFA varied widely. Lower-ranked teams showed stronger advantages at home, likely due to greater reliance on familiar environments, while top teams were more consistent across settings. This context sensitivity suggests performance depends not only on team ability but also on game environment, raising potential concerns about the consistency assumption in causal inference. Our multivariate modeling further strengthened these findings by accounting for correlations among game metrics, revealing how performance shifts across multiple dimensions together. This joint analysis offers a richer, more comprehensive view of how venue impacts team behavior.

A key assumption in causal inference is consistency that treatment assignment (home vs. away) uniquely determines the potential outcome. However, if teams adjust their style of play based on context—such as opponent strength, league position, or crowd presence—the potential outcome under ``home'' or ``away'' may not be fixed. Our findings do not violate the consistency assumption but suggest the presence of context-dependent treatment effects, particularly among teams with variable performance. Future work could address this by explicitly modeling effect heterogeneity or conducting sensitivity analyses to assess robustness to such violations. Looking ahead, several directions merit further exploration. First, decomposing HFA into specific causal mechanisms such as crowd support, travel demands, or field familiarity would enhance interpretability. Second, incorporating mediation analysis could clarify how in-game metrics influence final outcomes like wins or goals. Third, extending our model to allow for distribution-free inference would increase its applicability beyond settings that satisfy normality assumptions. Finally, applying this framework to natural experiments such as matches played without fans during the COVID-19 pandemic could help isolate the impact of external factors on home field advantage.


\appendix
\renewcommand\thetheorem{\Alph{section}.\arabic{theorem}}
\numberwithin{figure}{section}
\numberwithin{table}{section}

\renewcommand{\thesection}{\Alph{section}}

\newpage

\section{Univariate simulation results for estimated $\Delta$}
Table~\ref{tab:delta_results} summarizes the performance of
the estimated $\Delta$ in various data generation scenarios. The bias, coverage probability, sample variance of the 1,000 estimates and mean of the 1,000 variance estimates are reported.

\begin{table}[H]

    \centering
     \caption{Summary of simulation results for $\widehat{\Delta}$. (Column (1): mean bias, (2): coverage probabilities for estimators, (3): sample variance of the 1,000 estimates, and (4): mean of the 1,000 variance estimates).}\label{tab:delta_results}
     \renewcommand{\arraystretch}{0.6}
    \begin{tabular}{cclcccc}
    \toprule
      &  & & (1) Bias & (2) CP & (3) SV & (4) MV\\
     \midrule
  Scenario 1 &	$n=10$& $\sigma_0^2=0.5$&-0.0016&0.899&0.0143&0.0143\\
    &     & $\sigma_0^2=1$&-0.0042&0.913&0.0195&0.0198\\
    &            & $\sigma_0^2=2$&0.0073&0.893&0.0323&0.0345\\
                   	  \midrule
   &	$n=20$& $\sigma_0^2=0.5$&-0.0006&0.923&0.0061&0.0055\\
  &       & $\sigma_0^2=1$&-0.0059&0.914&0.0075&0.0073\\
    &            & $\sigma_0^2=2$&0.0055&0.950&0.0092&0.0106\\
                    \midrule
   &	$n=40$& $\sigma_0^2=0.5$&0.0002&0.928&0.0027&0.0025\\
   &      & $\sigma_0^2=1$&0.0023&0.941&0.0030&0.0030\\
   &             & $\sigma_0^2=2$&-0.0002&0.936&0.0039&0.0038\\
  \midrule
      &	$n=80$& $\sigma_0^2=0.5$&-0.0013&0.947&0.0012&0.0012\\
   &      & $\sigma_0^2=1$&0.0004&0.949&0.0013&0.0013\\
   &             & $\sigma_0^2=2$&0.0016&0.947&0.0015&0.0015\\
         \midrule
         Scenario 2 &	$n=10$& $\sigma_0^2=0.5$&0.0036&0.882&0.0149&0.0139\\
    &     & $\sigma_0^2=1$&0.0032&0.910&0.0192&0.0208\\
    &            & $\sigma_0^2=2$&-0.0004&0.913&0.0321&0.0344\\
                    \midrule
   &	$n=20$& $\sigma_0^2=0.5$&0.0011&0.927&0.0057&0.0055\\
  &       & $\sigma_0^2=1$&0.0003&0.928&0.0074&0.0072\\
    &            & $\sigma_0^2=2$&0.0016&0.933&0.0100&0.0107\\
                    \midrule
   &	$n=40$& $\sigma_0^2=0.5$&0.0019&0.949&0.0025&0.0025\\
   &      & $\sigma_0^2=1$&0.0008&0.940&0.0028&0.0029\\
   &             & $\sigma_0^2=2$&0.0027&0.945&0.0036&0.0038\\
    \midrule
      &	$n=80$& $\sigma_0^2=0.5$&-0.0009&0.954&0.0018&0.0018\\
   &      & $\sigma_0^2=1$&0.0004&0.949&0.0013&0.0013\\
   &             & $\sigma_0^2=2$&-0.0010&0.955&0.0014&0.0015\\
   \bottomrule
    \end{tabular}
   
\end{table}

\section{Univariate simulation results: Bias box plots of estimated $\mathcal{\beta}$}
The following figures (Figures~\ref{fig:betaBias20} to~\ref{fig:betaBias80}) summarize the bias boxplots for all estimators $\bm{\beta}$ in different data generation processes to show the estimation performance in the uni-variate simulation scenario.

\begin{figure}[!h]
    \centering
    \includegraphics[width=0.95\textwidth]{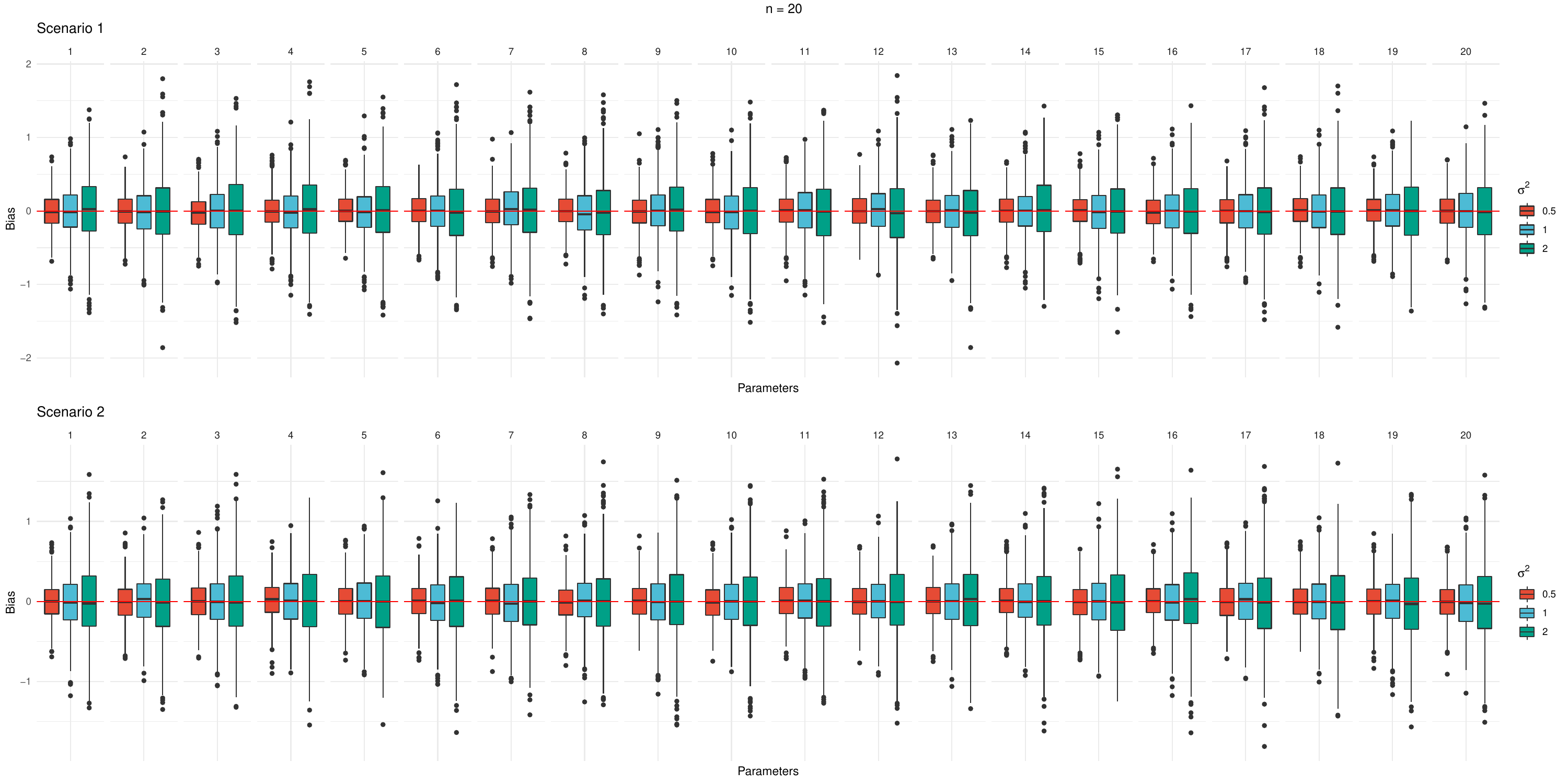}
    \caption{Simulation results: boxplot for Bias of $\widehat{\bm{\beta}}$ when $n=20$.}
    \label{fig:betaBias20}
\end{figure}

\begin{figure}[!h]
    \centering
    \includegraphics[width=0.95\textwidth]{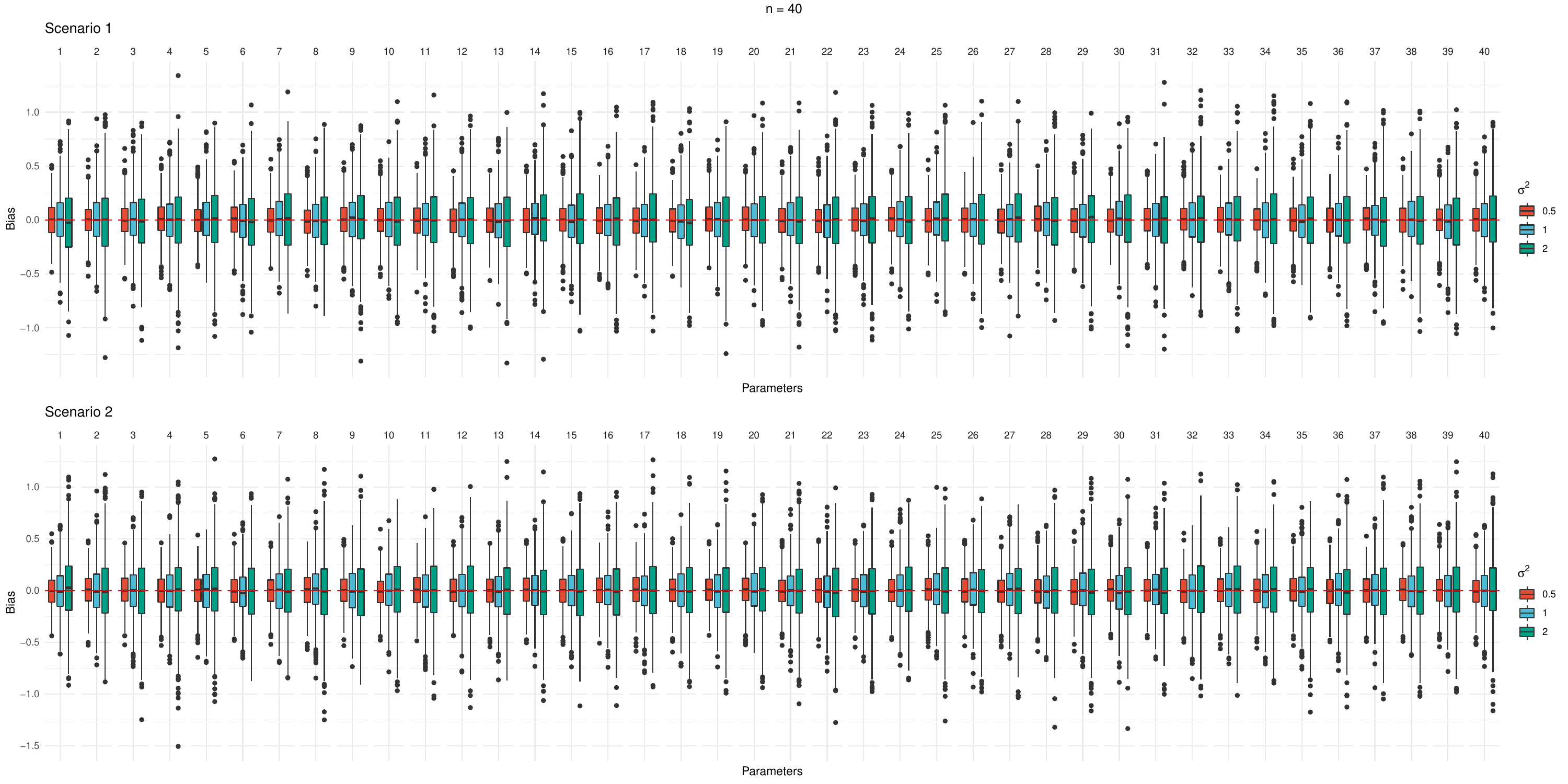}
    \caption{Simulation results: boxplot for Bias of $\widehat{\bm{\beta}}$ when $n=40$.}
    \label{fig:betaBias40}
\end{figure}

\begin{figure}[!h]
    \centering
    \includegraphics[width=0.95\textwidth]{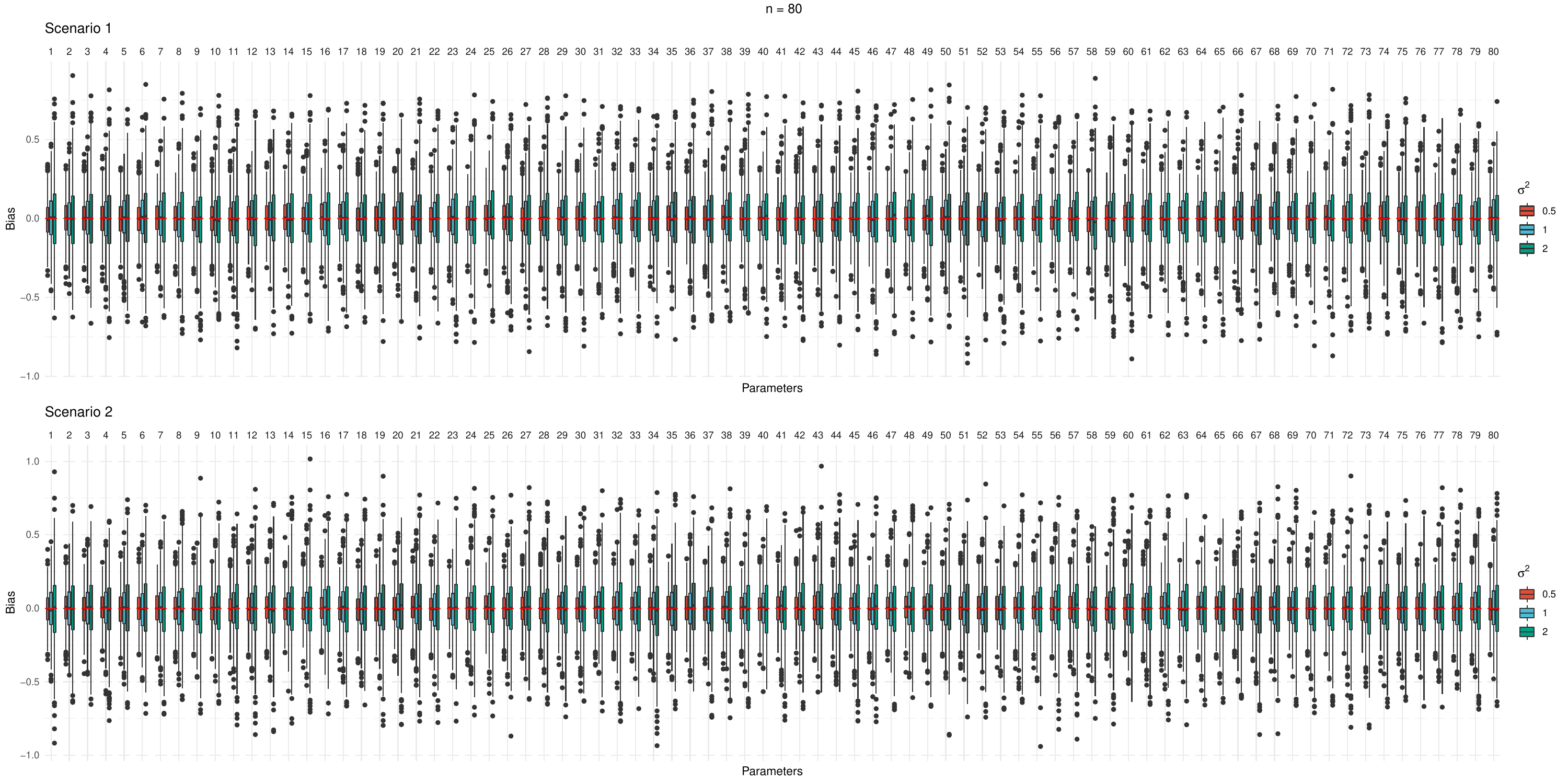}
    \caption{Simulation results: boxplot for Bias of $\widehat{\bm{\beta}}$ when $n=80$.}
    \label{fig:betaBias80}
\end{figure}

\section{Univariate simulation results with interactions between strength and home field advantage}
We carry out a simulation study with ranked $\bm{\beta}$. First we fix $\Delta=1$, $n=20$ and $\epsilon_{ij}\overset{\text{i.i.d}}{\sim} \mathcal{N}(0,\sigma^2_0), i=1,\ldots,n,\, j=1,\ldots,n$. We generate $\beta_i\overset{\text{i.i.d}}{\sim} \mathcal{N}(1,0.3^2), i=1,\ldots,n$, and rank them in a descending order. For each ranked team, we generate the ability $\text{Ab}_{i}\sim \mathcal{N}(0,2^2),i=1,\ldots,n$, and then set $\alpha_{ij}=\text{Ab}_{i}-\text{Ab}_{j}$ for $i=1,\ldots,n,\, j=1,\ldots,n$. We choose three different values for the variance of $\epsilon_{ij}$, $\sigma^2_0=0.5,1,2$.

\begin{figure}[!h]
    \centering
    \includegraphics[width=1.1\textwidth]{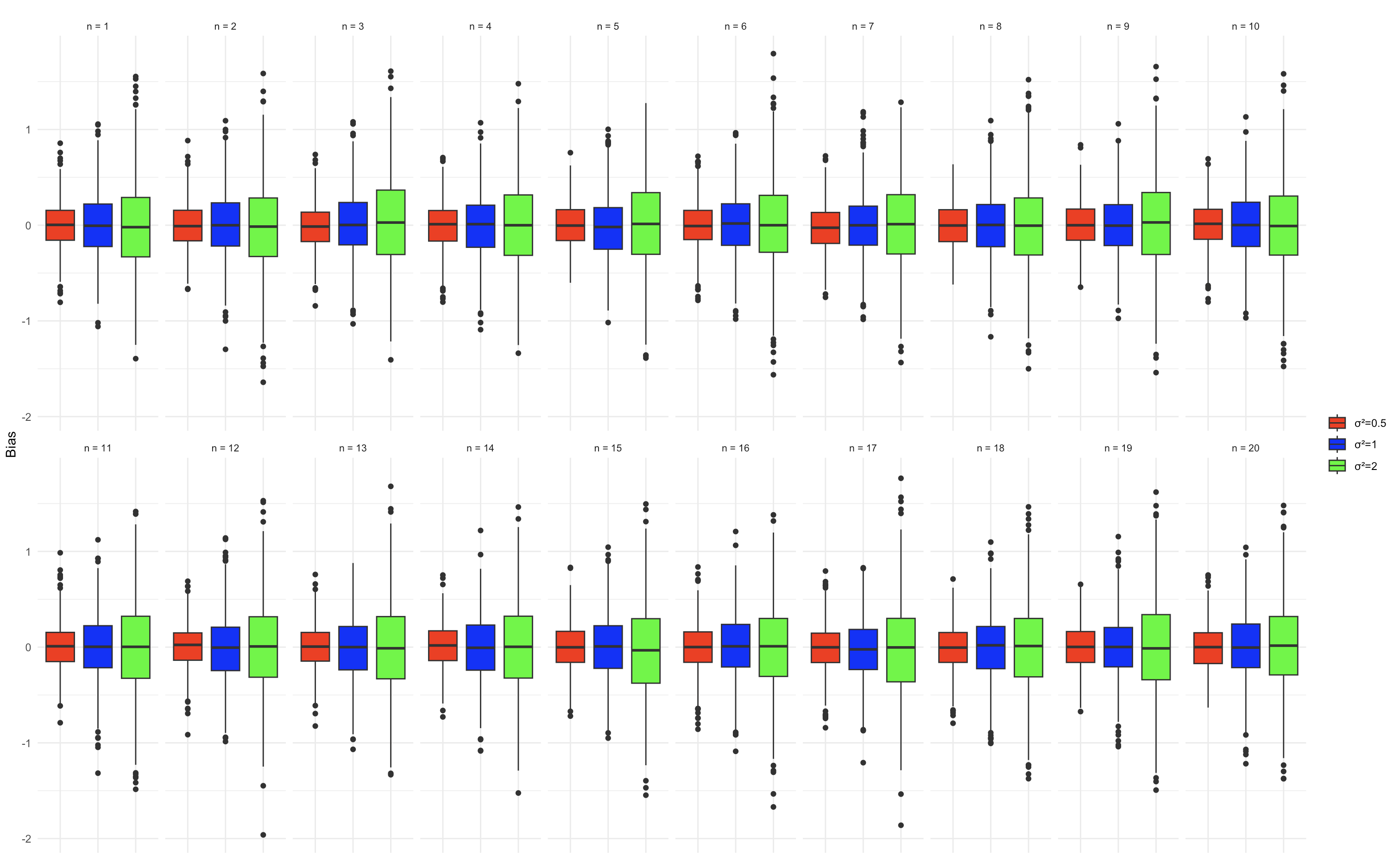}
    \caption{Simulation results: boxplot for bias of $\widehat{\bm{\beta}}$ grouped by $\sigma^{2}$ when $n=20$.}
    \label{fig:beta_biassimurank20}
\end{figure}

\begin{figure}[!h]
    \centering
    \includegraphics[width=0.8\textwidth]{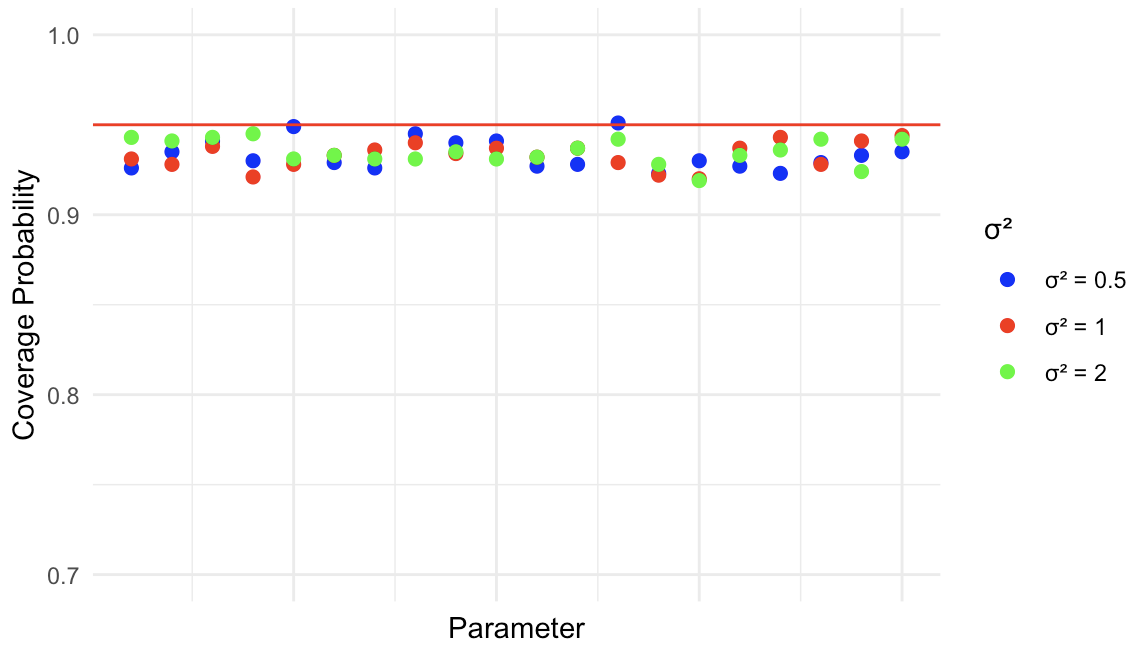}
    \caption{Simulation results: coverage probabilities for $\widehat{\bm{\beta}}$ grouped by $\sigma^{2}$ when $n=20$.}
    \label{fig:beta_cpsimurank20}
\end{figure}

Through 1,000 independent replicates for each scenario, the performance of the estimates are evaluated by the bias, the coverage probability (CP), the sample variance (SV), and the mean of the 1,000 variance estimates (MV). Table~\ref{supp:delta_results_simurank} summarizes the performance of the estimated $\Delta$. It is clear that our estimator for $\Delta$ has a small bias and the coverage probability close to the nominal coverage level of 95\%. From Figure~\ref{fig:beta_biassimurank20} of bias boxplots for $\widehat{\bm{\beta}}$, the estimator is essentially unbiased. Furthermore, all the coverage probabilities of our estimated $\bm{\beta}$ are fairly close to the nominal level of 0.95 in Figure~\ref{fig:beta_cpsimurank20}. In summary, the performance of our proposed model is invariant with interactions between strength and home field advantage.

\begin{table}[H]
    \centering
     \caption{Summary of simulation results for $\widehat{\Delta}$. (Column (1): mean bias, (2): coverage probabilities for estimators, (3): sample variance of the 1,000 estimates, and (4): mean of the 1,000 variance estimates).}\label{supp:delta_results_simurank}
    \begin{tabular}{clcccc}
    \toprule
   & & (1) Bias & (2) CP & (3) SV & (4) MV\\
     \midrule
	$n=20$& $\sigma_0^2=0.5$&-0.00288&0.929&0.0056&0.0055\\
    & $\sigma_0^2=1$&0.00113&0.942&0.0067&0.0072\\
         & $\sigma_0^2=2$&0.00107&0.954&0.0096&0.0108\\         	 
   \bottomrule
    \end{tabular}
\end{table}

\section{Correlation of $vec(\widehat{\mathcal{B}}^\top)$ in joint analysis of EPL data}
\begin{figure}
    \centering
    \includegraphics[width=1\linewidth]{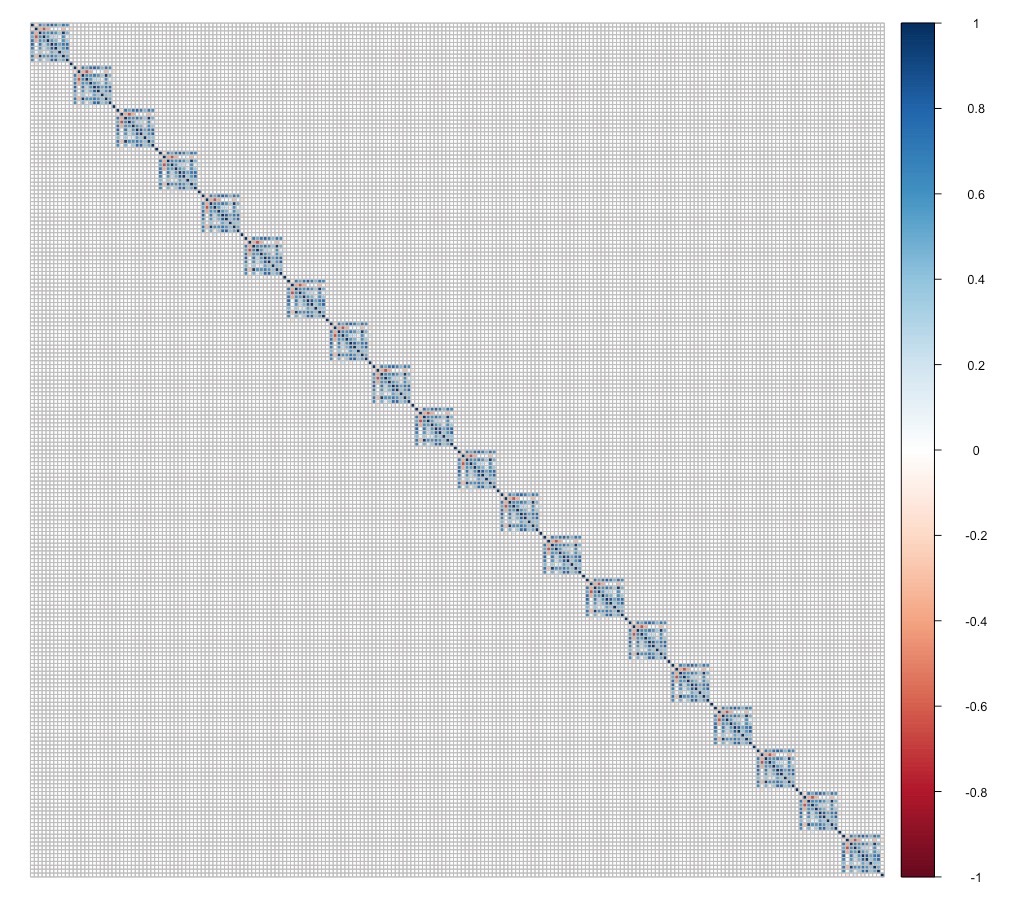}
    \caption{Correlation of $vec(\widehat{\mathcal{B}}^\top)$.}
    \label{fig:cor_B}
\end{figure}

In Figure~\ref{fig:cor_B}, we investigate the correlation of $vec(\widehat{\mathcal{B}}^\top)$, where the 20 $11*11$ block diagonals represent the correlations of 11 in-game statistics for each of 20 teams.

\section{Simulation study: Multivariate response case when $K=11$}
In this section, simulation studies are conducted in the multivariate response case when $K=11$. We fix  $\mathcal{D}=\mathbf{1}_k$ for all the simulation studies, $\mathcal{B}_i\overset{\text{i.i.d}}{\sim} \mathcal{N}(\mathcal{D},\Sigma_b)$, and $\mathcal{E}_i\overset{\text{i.i.d}}{\sim} \mathcal{N}(\mathbf{0}_k,\Sigma_E), i=1,\ldots,N/2$, where $\Sigma_b$ and $\Sigma_E$ follow an AR structure:
\[
\Sigma_{ij} = \rho^{|i-j|}, \quad \rho = 0.3, \quad i,j = 1,\dots,k.
\] We choose two different values for the number of teams, $n=60, 80$. In total, we implement $1,000$ independent replicates for each scenario.

Table \ref{tab:bias_D_hat_multi_simu} summarizes the performance of the estimator $\mathcal{D}$, and reports the bias and mean bias in all 11 dimensions. It clearly states that our proposed estimator for $\mathcal{D}$ has a small bias. In Table \ref{tab:cp_D_hat_multi_simu}, the coverage probabilities are also close to the nominal level of 0.95. From the table, the mean coverage probabilities approach the nominal coverage level 95\% as $n$ increases.

\begin{table}[ht]
\centering
\caption{Summary of simulation results for $\widehat{\mathcal{D}}$: bias of 11 dimensions and the mean bias}\label{tab:bias_D_hat_multi_simu}
\begin{tabular}{ccc}
  \toprule
  & Bias & Mean Bias \\ 
  \midrule
 n=60 & \makecell[l]{(0.00318, 0.00076, 0.00083, 0.00068, -0.00469, 0.00025,\\ 0.00148, 0.00029, -0.00387, -0.00319, -0.00005)} & -0.00039 \\ 
 n=80 & \makecell[l]{(0.00425, -0.00459, 0.00252, 0.00839, 0.00551, 0.00293,\\ 0.00269, -0.00038, 0.00070, 0.00220, -0.00431)} & 0.00181 \\ 
   \bottomrule
\end{tabular}
\end{table}

\begin{table}[ht]
\centering
\caption{Summary of simulation results for $\widehat{\mathcal{D}}$: coverage probability of 11 dimensions and the mean coverage probability}\label{tab:cp_D_hat_multi_simu}
\begin{tabular}{ccc}
  \toprule
 & CP & Mean CP \\ 
  \midrule
 n=60 & (0.969, 0.964, 0.968, 0.960, 0.976, 0.963, 0.955, 0.970, 0.969, 0.966, 0.969) & 0.966 \\ 
 n=80 & (0.957, 0.961, 0.961, 0.967, 0.964, 0.963, 0.971, 0.972, 0.961, 0.945, 0.951) & 0.961 \\ 
   \bottomrule
\end{tabular}
\end{table}

We examine the simulation results of $\widehat{\mathcal{B}}$ in multivariate response cases when $K=11$. We pick the first 3 teams in Table \ref{tab:bias_B_results_multivar} and ~\ref{tab:cp_B_results_multivar}, showing the bias, coverage probability, and their means across all 11 dimensions. Our estimator $\widehat{\mathcal{B}}$ shows a small bias. Furthermore, all coverage probabilities are close to the desired nominal level 95\%.

\begin{table}[ht]
\centering
\caption{Simulation Results for $\widehat{\mathcal{B}}$ with first 3 teams: Bias}\label{tab:bias_B_results_multivar}
\begin{tabular}{clcc}
  \toprule
&  & Bias & Mean Bias \\ 
  \midrule
 n=60 & Team 1 & \makecell[l]{(0.004933, -0.000604, -0.003024, 0.001300, -0.002927, 0.005478,\\ -0.006293, -0.000865, -0.003510, 0.000113, -0.000738)} & -0.000558 \\ 
  & Team 2 & \makecell[l]{(0.004071, -0.006439, 0.001129, 0.000972, -0.006614, -0.002784,\\ -0.004241, 0.004996, 0.005084, -0.003777, -0.007714)} & -0.001393 \\ 
  & Team 3 & \makecell[l]{(0.007058, 0.008165, 0.006222, 0.003592, 0.004475, 0.003697,\\ 0.002099, 0.008392, 0.001912, 0.002178, 0.001669)} & 0.004496 \\ 
 n=80 & Team 1 & \makecell[l]{(-0.006400, 0.001645, -0.003829, 0.000415, 0.003569, -0.004925,\\ -0.004529, -0.002377, -0.002875, 0.000944, 0.006578)} & -0.001071 \\ 
  & Team 2 & \makecell[l]{(0.002441, 0.001653, -0.001590, -0.005972, -0.002743, -0.006439,\\ -0.002146, 0.002233, -0.003476, -0.003870, -0.007925)} & -0.002530 \\ 
  & Team 3 & \makecell[l]{(0.006291, 0.007684, -0.001298, 0.004376, -0.001689, -0.003639,\\ 0.005016, -0.000351, 0.005410, -0.000117, -0.002894)} & 0.001708 \\ 
   \bottomrule
\end{tabular}
\end{table}

\begin{table}[ht]
\centering
\caption{Simulation Results for $\widehat{\mathcal{B}}$ with first 3 teams: Coverage Probability}\label{tab:cp_B_results_multivar}
\begin{tabular}{clcc}
  \toprule
&  & CP & Mean CP \\ 
  \midrule
 n=60 & Team 1 & \makecell[l]{(0.959, 0.956, 0.953, 0.944, 0.956, 0.949, \\ 0.956, 0.958, 0.947, 0.952, 0.955)} & 0.953 \\ 
  & Team 2 & \makecell[l]{(0.959, 0.952, 0.952, 0.936, 0.949, 0.954, \\ 0.964, 0.948, 0.952, 0.954, 0.944)} & 0.951 \\ 
  & Team 3 & \makecell[l]{(0.951, 0.948, 0.951, 0.948, 0.950, 0.946, \\ 0.948, 0.945, 0.943, 0.950, 0.956)} & 0.949 \\ 
 n=80 & Team 1 & \makecell[l]{(0.957, 0.951, 0.938, 0.957, 0.941, 0.948, \\ 0.945, 0.954, 0.952, 0.940, 0.953)} & 0.949 \\ 
  & Team 2 & \makecell[l]{(0.944, 0.952, 0.956, 0.947, 0.947, 0.953, \\ 0.942, 0.955, 0.962, 0.948, 0.948)} & 0.950 \\ 
  & Team 3 & \makecell[l]{(0.957, 0.954, 0.957, 0.964, 0.956, 0.956, \\ 0.943, 0.951, 0.945, 0.959, 0.957)} & 0.954 \\ 
   \bottomrule
\end{tabular}
\end{table}

In conclusion, the simulation results when K=11 verify the excellent performance of our proposed estimation methods in multivariate response cases in terms of bias and coverage probability.

\section{Final standings of English Premier League 2020–2021 season}
In Table~\ref{tab:epl_rankings}, we list all 20 teams in order of their final ranks.

\begin{table}[h]
\centering
\caption{English Premier League 2020-2021 Final Standings}
\label{tab:epl_rankings}
\renewcommand{\arraystretch}{0.8}
\begin{tabular}{cc}
\toprule
\textbf{Rank} & \textbf{Team} \\
\midrule
1  & Manchester City \\
2  & Manchester United \\
3  & Liverpool \\
4  & Chelsea \\
5  & Leicester City \\
6  & West Ham United \\
7  & Tottenham Hotspur \\
8  & Arsenal \\
9  & Leeds United \\
10 & Everton \\
11 & Aston Villa \\
12 & Newcastle United \\
13 & Wolverhampton Wanderers \\
14 & Crystal Palace \\
15 & Southampton \\
16 & Brighton \\
17 & Burnley \\
18 & Fulham \\
19 & West Bromwich Albion \\
20 & Sheffield United \\
\bottomrule
\end{tabular}
\end{table}

  \bibliography{ref.bib}

\end{document}